\documentclass[usenatbib]{mn2e}
\usepackage{graphicx}

\title[Viscous time scale and Lindblad resonances in LMXBs]{X-ray variability,
viscous time scale and Lindblad resonances in LMXBs.}  
 
\author[M.Gilfanov \& V.Arefiev]
	{Marat Gilfanov$^{1,2}$ and Vadim Arefiev$^2$\\
        $^1$Max-Planck-Institut f\"ur Astrophysik, 85741 Garching
            b. M\"unchen, Germany\\
	$^2$Space Research Institute, Russian Academy of Sciences,
        Profsoyuznaya 84/32, 117997 Moscow, Russia}

\date{\today}

\pubyear{2005}

\begin{document}

\maketitle

\begin{abstract}
Based on RXTE/ASM and EXOSAT/ME data we studied 
X-ray variability of persistent LMXBs in the $\sim 10^{-8}-10^{-1}$ Hz
frequency range, aiming to detect features in their power density
spectra (PDS) associated with the viscous time scale of the accretion
disk $t_{\rm visc}$. 
As this is the longest intrinsic time scale of the disk, the
power density  of its $\dot{M}$ variations is expected to be
independent on the frequency at $f\la 1/t_{\rm visc}$.
Indeed, in the PDS of 11 sources out of 12  we found very low
frequency break, below which the spectra are nearly flat. At higher
frequencies they approximately follow the $P_\nu\propto \nu^{-1.3}$ law. 

The break frequency correlates very well with the binary orbital
frequency in a broad range of  binary periods 
$P_{\rm orb}\sim 12{\rm ~min}- 33.5 {\rm ~days}$, in accord with 
theoretical expectations for  the viscous time scale of the disk. 
However, the value of $f_{\rm break}/f_{\rm orb}$ is at least by an
order of magnitude larger than predicted by the standard disk theory. 
This suggests that a significant fraction of the accretion $\dot{M}$
occurs through the optically thin and hot coronal flow with the aspect
ratio of $H/R\ga 0.1$. 
The predicted parameters of this flow, 
$T\sim 10^4-10^6$ K and $n_e\sim 10^{12}-10^{15}$ cm$^{-3}$
are in qualitative agreement with recent Chandra and XMM-Newton
observations of complex absorption/emission features in the spectra of
LMXBs with high inclination angle. 

We find a clear dichotomy in the value of $t_{\rm visc}/P_{\rm orb}$
between wide and compact systems, the compact systems having $\sim 10$
times shorter viscous time. The boundary lies at the mass ration
$q\approx 0.3$, suggesting that this dichotomy is caused by the
excitation of the 3:1 inner Lindblad resonance in low-$q$ LMXBs.  

\end{abstract}


\section{Introduction} 
\label{sec:intro}
 
\begin{figure}
\centering
\includegraphics[width=0.51\textwidth]{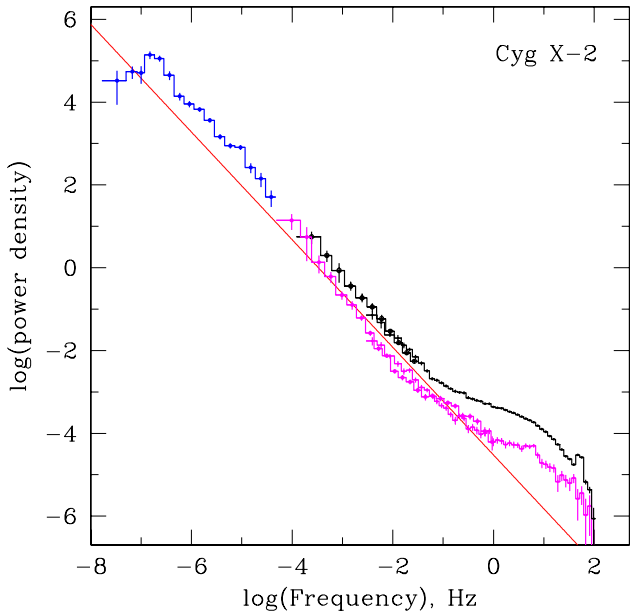}
\vspace{0.5cm}
\caption{The power spectrum of Cyg X-2 in broad frequency range. 
Based on non-simultaneous ASM, PCA and EXOSAT data. The solid line shows 
$P_\nu=1\cdot 10^{-4}\, \nu^{-1.3}$  power law.}
\label{fig:pds_cygx2}
\end{figure}

X-ray binaries show X-ray flux  variations  in a broad
range of time   scales (Fig.\ref{fig:pds_cygx2}). 
Their power density spectra often reveal a
number of periodic and quasi-periodic phenomena, corresponding to the
orbital  frequency of the binary system, the spin frequency of the
neutron star and  quasi-periodic oscillation, which nature is  still
poorly understood. In addition, aperiodic variability is observed,
giving rise to the broad band continuum component in  power density
spectra, extending from $\sim 10-100$ msec to the longest time scales
accessible for monitoring instruments.

The aperiodic variability is often characterized by the fractional
rms values of $\sim$ tens of per cent, indicating that variations of
the mass accretion rate $\dot{M}$ in the broad range of time scales
are present in  the innermost region of the accretion  flow, where
X-ray emission is produced. From the point of view of characteristic
time scales,  the high frequency  variations could potentially be
produced in the vicinity of the compact object. Longer time scales, on
the contrary,  exceed by many orders of magnitude 
the characteristic time scales in the region of the main energy
release. Hence, the low frequency $\dot{M}$ variations have to be
generated in the outer parts of the accretion flow and to be
propagated to the region of the main energy release.
Due to the diffusive nature of the standard Shakura-Sunyaev disk
\citep{ss73}, it plays the role of the low-pass filter, at any given
radius $r$ suppressing $\dot{M}$ variations on the time scales shorter
than the local viscous  time $t_{\rm visc}(r)$. The viscous time is, 
on the other hand, the longest  time scale of the
accretion flow. Hence, it is plausible to suppose that any given
radius $r$  contributes to  $\dot{M}$ and X-ray flux variations
predominantly at the frequency $f\sim 1/t_{\rm visc}(r)$
\citep{lyub97,chur01}.
\citet{lyub97} demonstrated that in this picture 
$P_{\nu}\propto \nu^{-\alpha}$  power density spectra can naturally
appear with slope $\alpha\sim 1$, in good agreement with
observations.

Owing to the finite size of the accretion disk, the longest time scale
which can appear in the disk is restricted by the viscous time on its
outer boundary  $t_{\rm visc}(R_d)$. Below this frequency the
X-ray flux variations are uncorrelated, therefore the power density
spectrum of the accretion disk should become flat at 
$f\la f_{\rm visc}\sim 1/t_{\rm visc}(R_d)$.\footnote{
The broad QPO like features which can appear near $t_{\rm visc}^{-1}$ 
due to the mechanism suggested by \citet{vikhl94} are discussed in
section  \ref{sec:qpo}} 
If there are several components in the  accretion flow each having a
different viscous time scale (e.g. geometrically thin disk and diffuse
corona above it), several breaks can appear on the power spectra at
the frequencies, corresponding to the inverse viscous time scale of
each  component.
Yet another source of $\dot{M}$ variations might be the variability of
the mass transfer rate from the donor star. As these variations also
have to be propagated through the  accretion disk, only low frequency 
perturbations, with $f\la 1/t_{\rm visc}(R_d)$, will reach the region of
the main energy release. This might give rise to an independent
continuum component in the power density spectrum at 
$f\la 1/t_{\rm visc}(R_d)$. 
The shape of the final power density spectrum of X-ray binary
will depend on the relative amplitudes of 
the $\dot{M}$ variations due to different components of the  accretion
flow and the donor star. As these are independent from each other, a
distinct features
should appear in the power density spectrum of X-ray binary at
frequencies $f\sim 1/t_{\rm visc}(R_d)$.

The viscous time scale of the  accretion disk is
\citep{ss73} 
\begin{equation}
t_{\rm visc}=\frac{2}{3\alpha}\,\left(\frac{H_d}{R_d}\right)^{-2}
\frac{1}{\Omega_K(R_d)}
\end{equation}
where $H_d$ is the disk half thickness and $\Omega_K(R_d)$ is the
Keplerian frequency at the outer disk boundary $R_d$. 
For a geometrically thin disk, $H/R\ll 1$, the viscous time scale is
significantly longer than the Keplerian time at the outer disk
boundary.  
If the disk fills a large fraction of the Roche-lobe of the primary,  as
is the case  for steady disks in persistent LMXBs,  
$R_d\sim 0.7-0.9\,R_L$, (section \ref{sec:tvisc_h2r}), the viscous
time at the outer boundary  $t_{\rm visc}(R_d)$ will exceed  the
orbital period of the binary system $P_{orb}$ as well. 
Therefore, one might expect to find  features associated with the
viscous time of the accretion disk at frequencies  below the
orbital frequency of the binary system. Detection of such features for
a  sample of LMXBs would give an opportunity to probe the
accretion disk parameters, such as viscosity parameter $\alpha$ and the
disk thickness at its outer boundary. 

In this paper we study a sample of persistent low mass X-ray binaries
aiming  to detect such features in their power density spectra.
Longevity of typical orbital periods in LMXBs requires 
long monitoring observations to achieve this goal.  
This has become possible thanks to the long operations of the All Sky
Monitor aboard Rossi X-ray Timing Explorer, which provided long
term light curves of hundreds of galactic  sources covering period of
$\sim 7-8$ years.  Some sources with rather short orbital period,
$\la$~few hours may be studied based on the EXOSAT data.

\begin{table*}
\renewcommand{\arraystretch}{1.1}
\caption[]{The binary system parameters for LMXBs from the ASM and
EXOSAT samples.} 
\label{tab:general}
\begin{tabular}{lcccccccc} 
\hline\hline 
Source & $M_1$ & $M_2$ & $P_{\rm orb}$ & $a$ & q & $i$ & D & 
Ref.\\ 
 & $M_{\odot}$ & $M_{\odot}$ & hrs &  $10^{10}$ cm & & $\deg$ &  kpc
&  \\
 & (1) & (2) & (3) & (4) & (5) & (6) & (7) & (8) \\
\hline 
\multicolumn{9}{c}{\it ASM sample}\\ 
\hline 
GRS~1915+105 &  $14.0 \pm 4.4$  &  $0.81 \pm 0.53$  & $804 \pm 36$ 
	& $747 \pm  78$  & $0.058\pm 0.033$ & $66\degr \pm 2\degr$  
	&  $12.1 \pm 0.8$  & 1,2 \\ 
GX~13+1      & 1.4              &  $5\pm 1$ $^a$         & 592.8  
	& $461 \pm 25$ & $3.6\pm 0.9$ &  ---  
	& $7 \pm 1$        & 3 \\ 
Cir~X-1      & 1.4          &  $4 \pm 1$      & 398.4  
	& $334 \pm  21$  & $2.9\pm 0.8$ &  ---  
	& $5.5$&  4 \\ 
Cyg~X-2      & $1.78\pm0.23$&  $0.60\pm0.13$  & 236.3 
	& $180 \pm  7$  & $0.34\pm  0.04 $ & $62.5 \pm 4\degr$  
	& $7.2 \pm 1.1$ &  5 \\ 
GX~349+2     & 1.4          &  $1.5\pm 1$     & 22.5$\pm$0.1  
	& $40.0\pm 4.7$ & $1.07\pm  0.73 $ & ---  
	& $9.2$  &  6 \\
Sco~X-1      & 1.4      &  0.42           & 18.92 
	& $30.5 \pm 1.1$   & $0.30\pm  0.05 $ & $38\degr$ 
	& $2.8 \pm 0.3$    & 7 \\
\hline 
\multicolumn{9}{c}{\it EXOSAT sample}\\ 
\hline 
EXO~0748-676 & 1.4           & $0.45$ & $3.824$ 
	& $10.6\pm  0.4$  & $0.32\pm 0.06 $ & $\sim 75\degr$  
	& $7.6$   &  8\\ 
4U~1636-536  & $1.4$ & $ 0.36$ & $3.793$ 
	& $10.3 \pm  0.4$  & $0.26\pm  0.05 $ & --- 
	&  5.9        & 9 \\ 
4U~1323-619  & 1.4           & 0.26           &  2.93  
	& $8.53 \pm  0.35$  & $0.19\pm 0.03 $ & $\la 80\degr$  
	& $15 \pm 5$     & 9 \\ 
4U~1916-053  & 1.4           & $0.09\pm0.02$  &  0.83  
	& $3.55 \pm 0.16$   & $0.064\pm 0.017 $ & $\ga 70\degr$ 
	& 10.0       & 10 \\ 
1H~1627-673  & $1.4$& $0.08\pm0.01$ &  $0.690$ 
	& $3.13 \pm  0.14$  & $0.057\pm 0.011 $ &  $\la 8\degr$ 
	& 8.0        11,12\\ 
4U~1820-303  & $1.4$          & $0.07\pm0.01$  & $0.190$ 
	& $1.32 \pm 0.06$   & $0.050\pm 0.010 $ & --- 
	& 8.0        & 13 \\ 
\hline\hline 
\end{tabular}\medskip\\
\parbox{\textwidth}{
(1) -- the mass of the compact object, if the uncertainty is not given
$\sigma=0.2 M_\odot$ is assumed; 
(2) -- the mass of the secondary, if the uncertainty is not given,
$10\%$ relative error is assumed; 
(3) -- orbital period; 
(4) -- binary separation computed from the 3rd Kepler law; 
(5) -- mass ratio $q=M_1/M_2$. For GRS~1915+105 and Cyg~X-2 from refs
(1) and (5) respectively, for other sources computed from the masses
given in the columns (1) and (2) of the table.
(6) -- inclination; 
(7) -- source distance; 
(8) -- reference for binary system parameters: 
$^1$ Harlaftis \& Greiner 2004; 
$^2$ Fender et al. 1999;
$^3$ Bandyopadhyay et al. 1999; 
$^4$ Johnston et al. 1999;
$^5$ Orosz and Kuulkers 1999; 
$^6$ Wachter \& Margon 1996;  
$^7$ Steeghs and Casares 2002; 
$^8$ Parmar et al. 1986; 
$^{9}$ an estimate based on the empirical $P_{orb}-$mass relation from
Patterson, 1984; 
$^{10}$ Nelson et al. 1986; 
$^{11}$ Levine et al. 1988; 
$^{12}$ Chakrabarty 1998; 
$^{13}$ Rappaport et al. 1987;
$^a$ the companion mass estimate is very uncertain. 
}
\end{table*}

The paper is structured as follows. In section \ref{sec:data} we
describe the selection criteria and sample of LMXBs. Power density
spectra and their approximations are presented in the section
\ref{sec:pds}. The viscous time scale of the accretion disks as
obtained from the power spectra analysis and constrains on parameters
of the outer disk are discussed in the section \ref{sec:tvisc}. In
section \ref{sec:discussion} we consider the vertical structure of the
semi-thick accretion flow, derive parameters of the disk corona and compare 
these with other evidence of the existence of diffuse coronal flow in
LMXBs. Our results are summarized in section \ref{sec:summary}. In the
Appendix we describe the method used to calculate power density
spectra from the ASM light curves.

\begin{figure*}
\centering
\includegraphics[width=\textwidth]{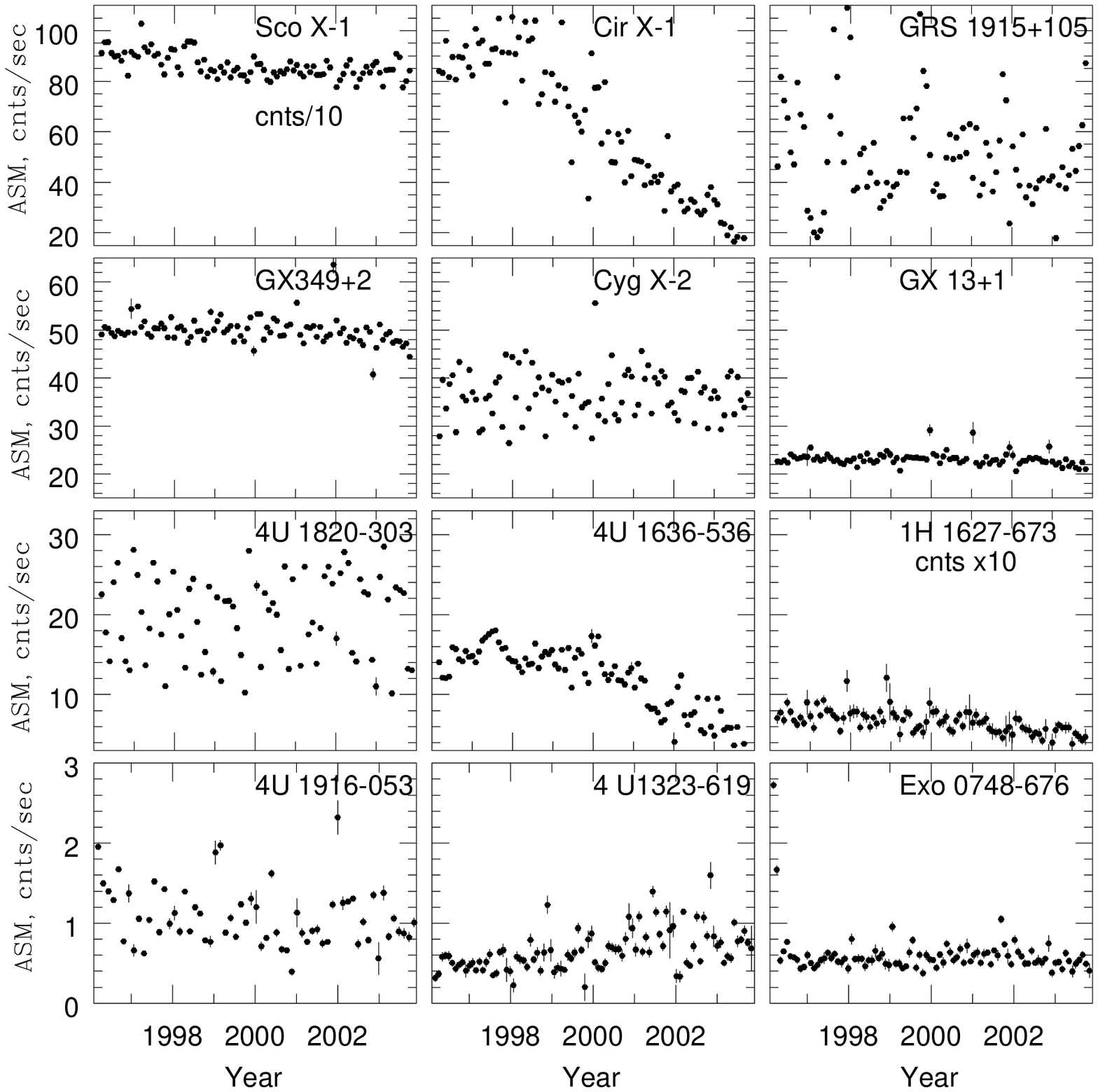}
\caption{Long term X-ray light curves of selected sources obtained
by ASM (2--12 keV). The light curves cover the period from 1996--2004}
\label{fig:lc_asm}
\end{figure*}

\section{The data and selection criteria} 
\label{sec:data} 

\subsection{Data}

We used data of the ASM instrument of RXTE observatory (Swank, 1999),
covering the period  from 1996--2004 (MJD range from 
$\approx 50100-53100$). The ASM instrument operates in the 2--12 keV energy
range and performs flux measurements for over $\sim 300$ X-ray
sources from the ASM source catalog once per satellite orbit,
i.e. every $\sim 90$~min. Each flux measurement (dwell) has duration
of $\sim 90$ sec. Due to navigational constrains and appearance of
very bright transient sources, the ASM  light curves for individual sources
sometimes have gaps of duration of up to few months.  The
dwell-by-dwell light  curves  were retrieved from the public RXTE/ASM 
archive at HEASARC.   
 
For sources with short orbital periods, $P_{\rm orb}\la 4$ hours,  we
used the data of the medium  energy (ME)  detector of EXOSAT satellite
\citep{turner81}. It provided $\sim$ several tens of ksec long
light curves with typical time resolution of $\sim 1$ sec in the
0.9--8.9 keV energy range.  The EXOSAT data were also retrieved from
HEASARC.

\subsection{Selection criteria} 
\label{sec:selection}
 
We have selected low mass X-ray binaries, satisfying the following
criteria:  
\begin{enumerate}
 
\item 
{\em Persistent sources}, which light curves do not show off-states or
X-ray Novae-like outbursts. This ensures that the accretion disk was
in the same state during the analyzed period.

\item
The {\em orbital period} $P_{orb}$ is known.  Orbital  parameters of
X-ray sources  were taken from the catalog of \citet{liu01}.

The Nyquist frequency in the power density spectra obtained from the
ASM light curves, $f_{\rm Nyq}\sim 10^{-4}$ Hz, correspond to period of
$\sim 3$ hours. In practice, due to specifics of the ASM light curves,
the high frequency part of the power spectra can sometimes be
distorted by the aliasing effect. To minimize the contribution of this
effect we selected source with sufficiently long orbital periods,  
$P_{\rm orb}\ga 10$ hours. 

Although the aliasing is not an issue for EXOSAT light curves, their duration,
typically limited by $\sim 50$ ksec, imposes a different constrains on
the $P_{\rm orb}$.
As we are looking for features at the time scales larger than 
$P_{\rm orb}$, we required that the light curve covers at least
several $P_{\rm orb}$. This leads to the constrain for the sources
which can be studied using EXOSAT light curves, 
$P_{orb} \la 4$ hours. 

\item 
{\em Source brightness.} We selected sources with the average count
rate exceeding $\ga 2$ cnts/s for ASM and $\ga  5$ cnts/s for EXOSAT.  
 
\item 
The {\em noise level}, calculated for ASM power spectra, although 
approximately correct, is not accurate enough, due to existence of
unaccounted systematic errors in the flux measurements
\cite[e.g.][]{grimm02}. 
To avoid additional uncertainties in the  power
spectra we   considered only sources with sufficiently high signal at
the orbital frequency, 
$P_{\rm signal} \ga 10  P_{\rm noise}$ at $f_{\rm orb}$.  
For EXOSAT light curves the noise level is not an issue.

\end{enumerate}

\subsection{The sample}

There are 6 sources from the ASM catalog and 6 sources observed by EXOSAT, 
satisfying the selection criteria of the section \ref{sec:selection}.
The final ASM sample includes the following sources: 
GRS~1915+105, GX~13+1, Cir~X-1, Cyg~X-2, GX~349+2, Sco~X-1 
The EXOSAT sample includes: 
4U~1323--619, 4U~1636--536, 4U~1820--303,  4U~1916--053, 1H~1627--673, 
Exo~0748--676

The binary system parameters for selected sources are listed in Table
\ref{tab:general}  and their X-ray light curves obtained by ASM are
shown in Fig.~\ref{fig:lc_asm}.

\section{Power density spectra} 
\label{sec:pds}

Power density spectra  were computed in the 2-12 keV (ASM) and
0.9--8.9 keV (EXOSAT ME) energy range. The PDS of the 
sources from ASM sample were obtained using the method based on the
autocorrelation function calculation as  described in
the Appendix \ref{sec:method}. The EXOSAT light curves were analyzed 
with the aid of the \textit{powspec} task from FTOOLS~5.1. In
analyzing the EXOSAT data we averaged the power spectra
obtained in (nearly) all individual observations with adequate time
resolution available in the public archive at HEASARC.

All but one source in the EXOSAT sample are X-ray bursters. The
presence of X-ray bursts in their light curves can results in
appearance of an additional component in the PDS, having no relation
to the power spectrum of $\dot{M}$ variations due to the accretion disk.  
To avoid this contamination, we screened the EXOSAT light curves to
exclude the time intervals corresponding to X-ray bursts. 
In the case of EXO 0748--676 we also excluded the time intervals
corresponding to the X-ray eclipses. No attempt to screen out the
X-ray dips has been done, due to ambiguity of their identification.
The power spectra based on the ASM data are not subject to a
significant contamination due to X-ray bursts and dips/eclipses.

\begin{figure*}
\vspace{2cm}
\centering
\includegraphics[width=\textwidth]{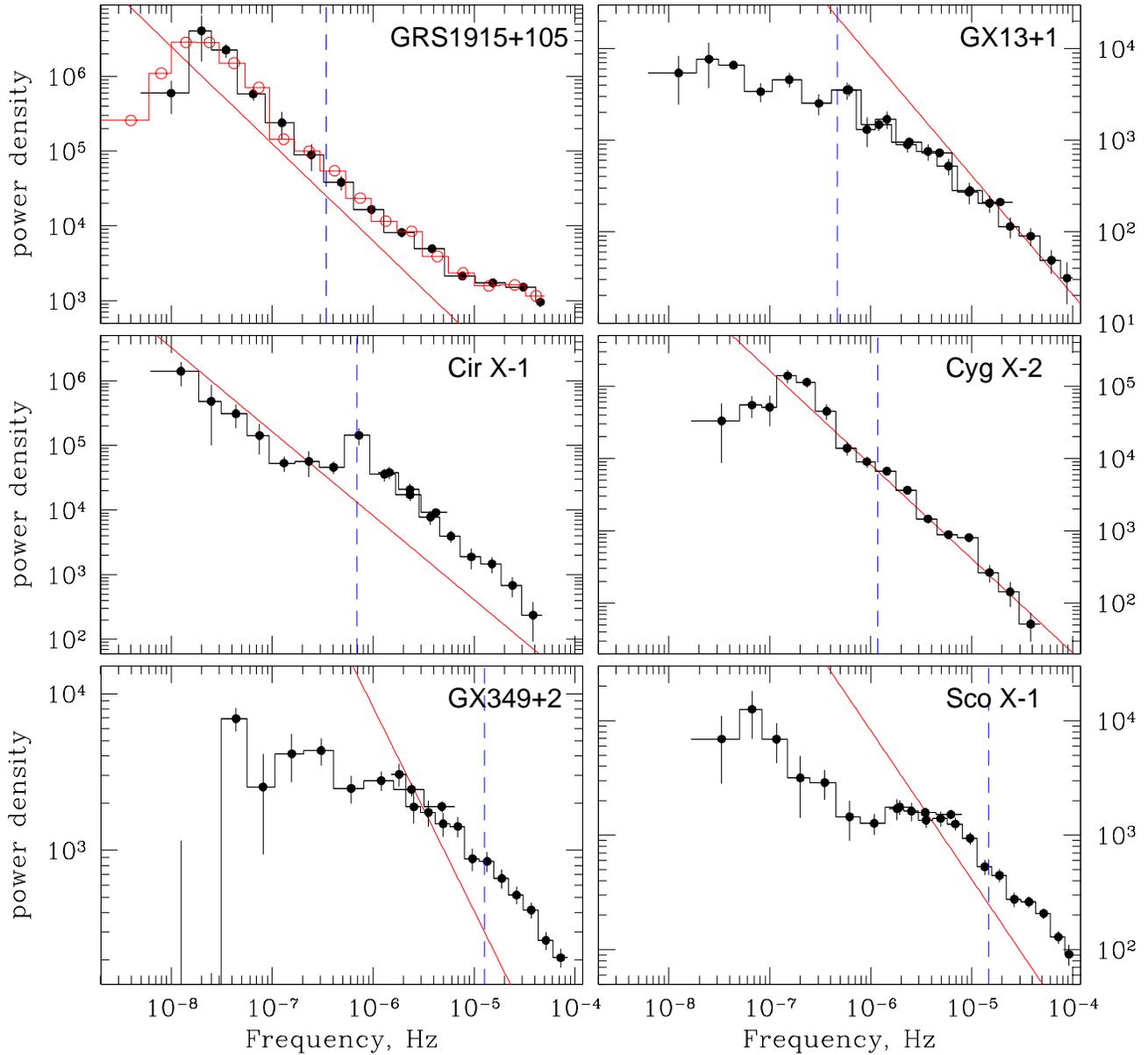}
\caption{Power density spectra of LMXBs from the ASM sample. The solid 
line in each panel shows $P_\nu=1\cdot 10^{-4}\, \nu^{-1.3}$  power
law. The vertical dashed line marks the orbital frequency. In the
upper left panel the histogram marked with hollow circles without error
bars shows the power spectrum of GRS1915+105 obtained from the
entire ASM time series of duration $\approx 8$ years. Note that
the vertical scale is different in different panels.} 
\vspace{1cm}
\label{fig:pds_asm}
\end{figure*}

\begin{figure*}
\vspace{2cm}
\centering
\includegraphics[width=\textwidth]{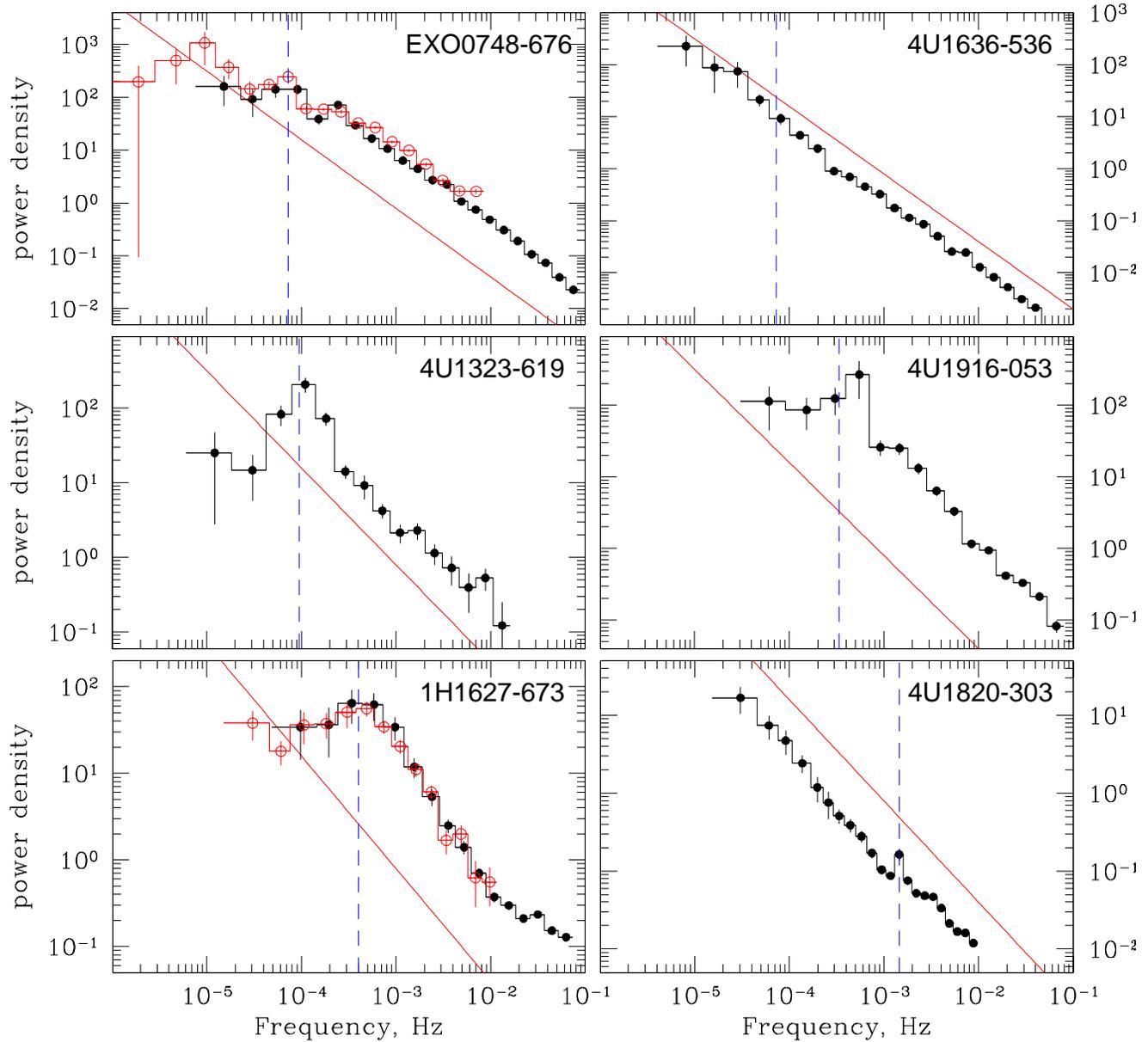}
\caption{Power density spectra of LMXBs from the EXOSAT sample
obtained from the ME data. The solid line in each panel shows
$P_\nu=1\cdot 10^{-4}\, \nu^{-1.3}$ 
power law, the same as in Fig.\ref{fig:pds_asm}. 
The vertical dashed line marks the orbital frequency.
For EXO0748--676 and 1H1627--673 the power spectra obtained from the
GSPC data are also shown (histograms marked with hollow circles). 
Note that the vertical scale is different in different panels.
}
\vspace{1cm}
\label{fig:pds_exosat}
\end{figure*}

\begin{figure*}
\centering
\includegraphics[width=0.94\textwidth, bb=100 200 480 700]{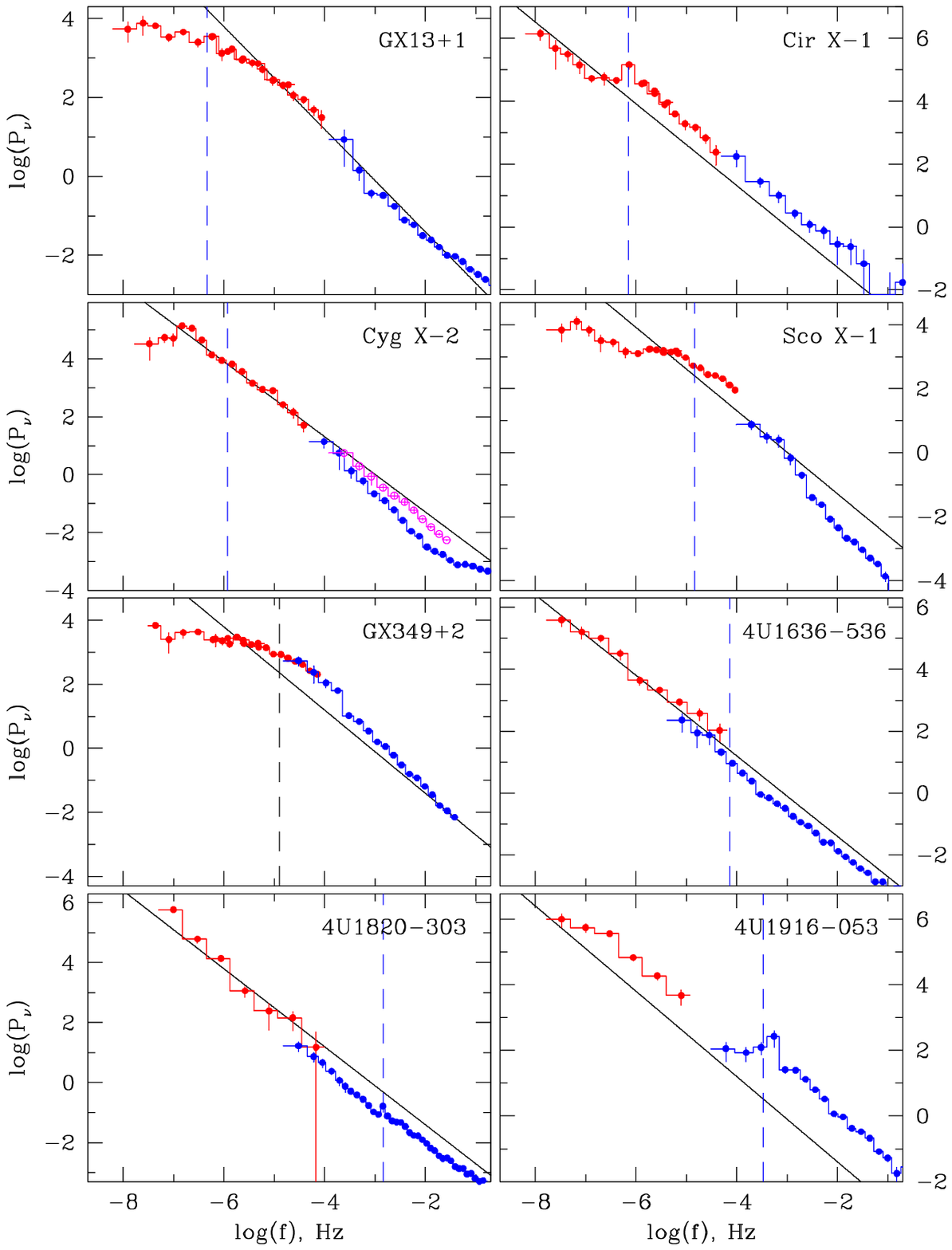}
\caption{Broad band power spectra of LMXBs based on ASM and EXOSAT data.
The solid line in each panel shows $P_\nu=1\cdot 10^{-4}\, \nu^{-1.3}$ 
power law, the same as in Fig.\ref{fig:pds_asm} and
\ref{fig:pds_exosat}. 
The vertical dashed line marks the orbital frequency.
For Cyg X-2 the power spectrum obtained from the RXTE/PCA data is also
shown (the histograms marked with open circles). 
Note that the vertical scale is different in different panels.
}
\label{fig:pds_broadband}
\end{figure*}

The  power spectra of selected sources are shown in
Figs.\ref{fig:pds_asm} and \ref{fig:pds_exosat}. 
For reference, we plot in each panel  a power law
spectrum  $P_\nu=1\cdot 10^{-4}\, \nu^{-1.3}$.
Flattening at frequencies $f\ga10^{-5}$ Hz can be seen in the PDS
of GRS~1915+105.
This can potentially be related to the specifics of ASM light curves,
namely, it can be caused by the aliasing effect leading to the leakage
of higher frequency power below the Nyquist frequency. 
The assessment of the reality of this flattening is beyond the scope of
the present paper.
Also seen for some sources (most prominently for Cir~X-1 and
4U1820--303) are the peaks due to the orbital  modulation. 
In Fig.\ref{fig:pds_broadband} we show combined  power spectra based on
the ASM and EXOSAT light curves. These spectra cover $\sim
10^{-8}-10^{-1}$ Hz frequency range. For several faint sources from
the EXOSAT sample no meaningful power spectra can be obtained from the
ASM light curves due to too low count rate.  
For GRS1915+105, obviously, no EXOSAT data exist.

The power spectra of almost all sources show a clear low
frequency break, below which they are nearly
flat. For some sources the second very low frequency component is
present at lower frequencies. The most clearly this component can be
seen in the case of Cir X-1, Sco X-1 and 4U1916--536. Some sources
also show broad QPO like features near or below the break frequency.   
At high frequencies, above the break, the power spectra follow the
$P_\nu\propto \nu^{-\alpha}$ power law. Remarkably, the slope of the
power law appears to be similar in all 12 sources,  
$\alpha\sim 1.3$. The same is true for the normalization.

\subsection {4U~1636-536 and 4U~1820-303}

There is at least one exception from the above behavior.
No evidence for break was found in the  PDS of 4U~1636-536. 
Although the power spectrum of this source might have several weak
features, its overall behavior in an extremely broad frequency range
from $10^{-8}-10^{-1}$ Hz is well represented by the power law with
slope $\alpha\approx 1.3$ and without any obvious break
(Fig.\ref{fig:pds_broadband}). 

\begin{figure}
\centering
\includegraphics[width=0.5\textwidth]{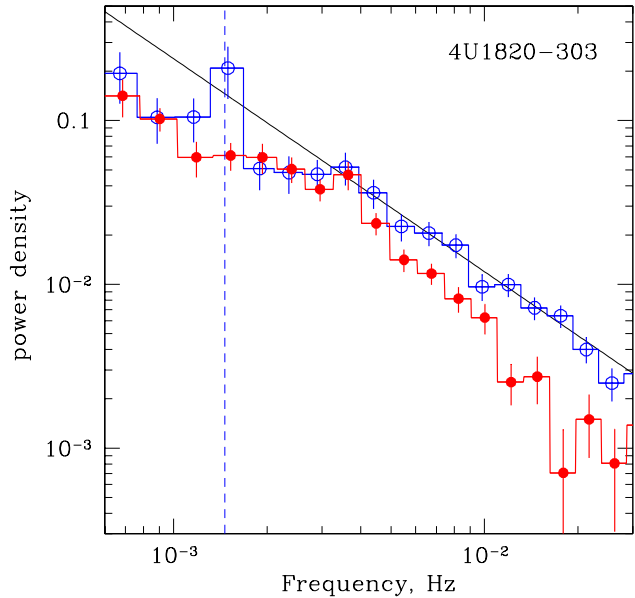}
\caption{The power spectrum of 4U1820-303 obtained by EXOSAT on April
16 (open circles) and August 19--20 (filled circles), 1985. 
The solid line shows $P_\nu=3\cdot 10^{-5}\, \nu^{-1.3}$ 
power law. The vertical dashed line marks the orbital frequency.}
\label{fig:pds_4u1820}
\end{figure}

The second possible exception is 4U~1820-303. 
In addition to clearly visible peak due to the orbital modulations, it 
appears to have a shoulder at $f\sim 10^{-3}-10^{-2}$ Hz followed by a
power law at lower frequencies. The presence of this shoulder is
more obvious in the data of individual EXOSAT observations, shown in 
Fig.\ref{fig:pds_4u1820}. 
In the formal statistical sense the existence of the low frequency
break is highly significant. For example, approximation of the
Aug. 19--20, 1985 data  in the
$10^{-3}-2\cdot 10^{-2}$ Hz frequency range gave the following values: 
$\chi^2=12.7$ (11 d.o.f.) and  $\chi^2=42.2$ (12
d.o.f.) for the power law model with and without low frequency
break respectively, resulting in $\Delta \chi^2\approx 30$ for one
additional parameter. However, considering the overall shape of the
power spectrum (e.g. Fig.\ref{fig:pds_broadband}),  this source
presents a less obvious case than the others in our sample.

\citet{chou01} analyzing long term X-ray variability of
4U~1820-303 confirmed earlier suggestion  that it is
a hierarchical triple system, where ultra-close X-ray binary is
orbited by companion with orbital period about 1.1 day. 
Due to influence of the third star  the well-known 176 day X-ray
modulation of  4U~1820-303 is generated.  Influence of the third star
can also give rise to very low frequency modulation of the mass
transfer rate in this system, unrelated to the $\dot{M}$ variations
intrinsic to the accretion disk.  This could, in principle, explain
strong very low  frequency power law component  at $f\la 10^{-3}$ Hz. 
In the following we associate the  shoulder at $\log(f)\sim -2.5$
with the low frequency break similar to the ones observed in other
LMXBs in our sample.  
We emphasize that this interpretation is not unique and 4U~1820-303,
like 4U~1636-536, can show the behavior different from other sources
in our sample.

\subsection{PDS approximation}

The power density spectra were approximated by a model of a power law
with low frequency break:
\begin{eqnarray}
\renewcommand{\arraystretch}{1.5}
P(f)=\left\{ \begin{array}{ll}
A\, f^{-\alpha} 	& \mbox{~~~ $f\ge f_{\rm break}$}\\
A\, f_{\rm break}^{-\alpha}& \mbox{~~~ $f< f_{\rm break}$}
\end{array} \right.
\label{eq:lfd}
\end{eqnarray}
An additional Lorentzian component was added to the model for the
sources with strong orbital modulation and/or broad low frequency QPO peaks.
The frequency range was chosen for each source individually, depending
on the duration of the available time series and the presence of the
additional noise component below $f_{\rm break}$.
The adopted frequency ranges and best fit values of the model
parameters are listed in Table 2, along with the count rates and X-ray
luminosities of the sources. The large width of the
QPO component present in the power spectra of several sources close to
the low frequency end makes the determination of the break frequency
somewhat ambiguous. For these sources (GRS1915+105, Cyg X-2 and
4U~1323--619) we list the values of the break frequency obtained with
and without the Lorentzian component in the model. In the following the
average of the frequencies of the two models are used. The statistical
uncertainty of the break frequency in these sources  are obtained from
combined confidence intervals of the models with and without Lorentzian
component.   
At least three harmonics of the orbital frequency are present in the
power spectrum of 4U1323--619. The corresponding bins of the power
spectrum were excluded from the fitting procedure.

\begin{table*}
\caption[]{The parameters of power spectra approximation.} 
\label{tab:pds}
\renewcommand{\arraystretch}{1.1}
\centering
\begin{tabular}{lccccccc} 
\hline\hline 
Source & ASM & EXOSAT & $L_X$ 
& $f_{\rm break}$ & $\alpha$    
& $f_{\rm QPO}$ & $f$ range \\ 
       & cnts/sec & cnts/sec & $10^{37}$ erg/s  & Hz    &  & Hz & Hz\\
& (1) & (2) & (3) & (4) & (5) & (6) & (7) \\
\hline 

GRS 1915+105 & 60.7 &--- & 34
&$4.03_{-0.64}^{+1.07}\times10^{-8}$&$1.32_{-0.09}^{+0.08}$
& --- & $ 5\cdot 10^{-9}- 1\cdot 10^{-6}$\\

             &&& &$6.50_{-1.64}^{+2.78}\times10^{-8}$&$1.26_{-0.11}^{+0.18}$  
&$2.8_{-2.8}^{+7.0}\times 10^{-8}$&$5\cdot 10^{-9}- 1\cdot 10^{-6} $\\

GX 13+1      & 23.1 &--- & 4.3
&$1.13\pm0.25\times10^{-7}$&$0.56\pm 0.03$
&---& $1\cdot 10^{-8}- 1\cdot 10^{-5}$\\

Cir X-1   & 74.2 &--- & 8.6
& $ 1.14_{-0.18}^{+0.12}\times 10^{-6}$& $1.51\pm 0.13$
& $7.26_{-0.84}^{+1.11}\times 10^{-7}$&$1\cdot 10^{-7}- 1\cdot 10^{-5}$\\

Cyg X-2   & 37.9 &--- & 7.5
& $ 2.28_{-0.38}^{+0.15}\times 10^{-7}$& $1.26\pm 0.06$&
--- &$1\cdot 10^{-8}- 1\cdot 10^{-5}$\\

          &&&& $ 3.66_{-1.47}^{+0.36}\times 10^{-7}$& $1.13\pm 0.09$
& $1.87\pm 0.09\times 10^{-7}$& $ 1\cdot 10^{-8}- 1\cdot 10^{-5}$\\

GX 349+2   & 51.6 &--- & 16.7
& $ 2.78_{-0.92}^{+0.63}\times 10^{-6}$& $0.66\pm 0.06$
&--- & $1\cdot 10^{-8}- 1\cdot 10^{-5}$\\

Sco X-1    & 917 &--- & 27.4
& $6.14_{-0.71}^{+1.06}\times10^{-6}$&$0.89_{-0.06}^{+0.10}$
&---& $ 5\cdot 10^{-7}- 5\cdot 10^{-5}$\\

EXO 0748--676   & 0.57 & 44.0 & 0.37
& $ 1.03_{-0.12}^{+0.16}\times 10^{-4}$&$1.25\pm 0.01$
&--- & $1\cdot 10^{-5}- 3\cdot 10^{-2}$\\

4U1636--536     & 14.7 & 153 &
&--- & $1.29\pm 0.01 $ 
& ---& $1\cdot 10^{-5}- 3\cdot 10^{-2}$ \\

4U1323--619     & 0.60 & 4.4 & 0.14
& $ 2.01\pm 0.22\times 10^{-4}$& $1.26_{-0.08}^{+0.10}$
&---& $1\cdot 10^{-5}- 3\cdot 10^{-2} $\\

              &&&& $ 2.10\pm 0.60\times 10^{-4}$& $1.21\pm 0.11$
& $ 8.14_{-0.30}^{+0.36}\times 10^{-5}$& $1\cdot 10^{-5}- 3\cdot 10^{-2} $\\

4U1916--053    & 1.0 & 17.8 & 0.26
& $ 4.25_{-0.83}^{+1.22}\times 10^{-4}$& $1.41\pm0.05$
&---&$3\cdot 10^{-5}- 3\cdot 10^{-2}$\\

1H1627--673    & 0.59 &44.5 & 0.41
& $ 5.78_{-1.25}^{+0.86}\times 10^{-4}$& $1.53\pm 0.08$
&---& $4\cdot 10^{-5}- 3\cdot 10^{-2}$\\

4U1820--303    & 19.8 &368 & 3.4
& $ 2.99_{-0.20}^{+0.21}\times 10^{-3}$& $1.43\pm 0.06$
&---& $4\cdot 10^{-5}- 3\cdot 10^{-2}$\\

\hline\hline 
\end{tabular}
\medskip\\
\parbox{\textwidth}{
(1), (2)  -- ASM and EXOSAT count rates; 
(3) -- X-ray luminosity calculated using respective  count rates
and source distances listed in Table \ref{tab:general};
(4) -- break frequency;
(5) -- power law slope;
(6) -- the centroid frequency of the Lorentzian component;
(7) -- frequency range for PDS fits.
}
\end{table*}

\section{Viscous time scale in LMXBs} 
\label{sec:tvisc}

For the standard accretion disk (Shakura \& Sunyaev, 1973) the viscous
time scale at the outer edge of the disk $R_d$ can be estimated as: 

\begin{equation}
t_{\rm visc}=\frac{2}{3\alpha}\,\left(\frac{H_d}{R_d}\right)^{-2}
\frac{1}{\Omega_K(R_d)}
\label{eq:tvisc}
\end{equation}
where $R_d$ is the disk outer radius,  
$\Omega_K = \sqrt{GM_1/R_d^3}$ is the Keplerian frequency, 
$H_d$ is the disk half-thickness at the outer edge, $M_1$ is the mass
of the primary and $\alpha$ is the dimensionless viscosity
parameter. 

Combining the eq.(\ref{eq:tvisc}) with the third Kepler law:
\begin{equation}
P_{orb} = 2 \pi a^{3/2} [G(M_1+M_2)]^{-1/2},
\label{eq:kepler3}
\end{equation}
and assuming that the binary has a circular orbit one finds:
\begin{equation}
\frac{f_{\rm visc}}{f_{\rm orb}} = \frac{3\pi \alpha}{\sqrt{1+q}} 
              \left(\frac{H_d}{R_d}\right)^2 
               \left(\frac{R_d}{a}\right)^{-3/2}
\label{eq:fvisc2forb}
\end{equation}
where $q=M_2/M_1$ is the mass ratio, $a$ is binary separation,
$f_{\rm orb}=1/P_{\rm orb}$ and $f_{\rm visc}=1/t_{\rm visc}$. 
From eq.(\ref{eq:fvisc2forb}) it follows that for fixed viscosity
parameter and relative disk thickness and size, the viscous time scale
is directly proportional to the orbital frequency of the binary
system, as intuitively expected.  

To estimate the expected range of values of $f_{\rm visc}/f_{\rm orb}$
one needs to know the disk size and thickness.
The thickness of the standard Shakura--Sunyaev disk is
\begin{equation}
\frac{H}{R}\approx 1.9 \cdot 10^{-2} \ \alpha^{-1/10} \ 
R_{\rm 10 km}^{3/20} \  M_{1.4}^{-21/40} \ 
L_{37}^{3/20} \ R_{10}^{1/8}
\label{eq:h2r_ss73}
\end{equation}
where 
$M_1=1.4\,M_{1.4}$ is the mass of the compact object in solar units,
$R_0=10^6\,R_{\rm 10 km}$ cm is defined so that $L_X=GM_1\dot{M}/R_0$,
$L_{X}=10^{37}\ L_{37}$ erg/s is X-ray luminosity,
and  $R_d=10^{10}\ R_{10}$ cm is the outer disk radius.
The disk thickness predicted by this relation varies from 
$H/R\sim 0.02$ for the most compact and low luminosity systems to
$H/R\sim 0.03-0.04$ for sources in the upper part of the Table
\ref{tab:general}. Irradiation of the disk by the X-rays from the
vicinity of the compact object, although does increase significantly
the surface temperature of the outer disk, has a little effect on its
mid-plane temperature, due to large optical depth of the
Shakura-Sunyaev disk \citep{lyutyi76}. Correspondingly, account for
irradiation effects  does not change the above estimates
significantly.  
With plausible values of two other parameters in eq.(\ref{eq:fvisc2forb}),
$\alpha\approx 0.5$ and $R_d/a\approx 0.3-0.6$, 
the standard theory  predicts:
\begin{equation}
\left( \frac{f_{\rm visc}}{f_{\rm orb}}\right)_{\rm std}\sim
0.005-0.07
\label{eq:fvisc2vorb_std}
\end{equation}

\begin{figure*}
\centering
\includegraphics[width=0.7\hsize, bb=18 266 592 670]{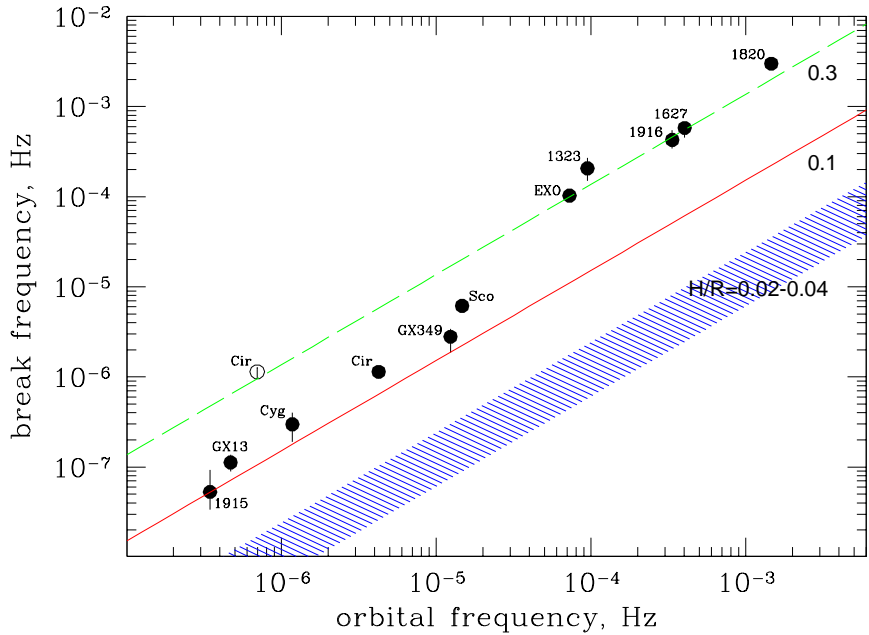}
\caption{Relation between the PDS break frequency and the
orbital frequency of the binary system. The shaded area 
shows the  dependence $f_{\rm visc}$ vs. $f_{\rm orb}$ expected
for the Shakura--Sunyaev disk, assuming $\alpha=0.5$, $\beta=f_{\rm
break}/f_{\rm visc}=1$, $q=0.5$ and  $R_d/a=0.4$. 
Straight solid and dashed lines are predictions for larger values of
the disk thickness $H/R$, as indicated by the numbers on the plot.
The two points for Cir X-1 correspond to the observed orbital period
of $P_{\rm orb}\approx 16.6$ d (open circle) and corrected for the
orbital eccentricity assuming $e=0.7$, as described in the section
\ref{sec:cirx1} (filled circle). 
} 
\label{fig:fbreak_forb}
\end{figure*}

\subsection{Break frequency}

Power density spectra of all but one  LMXBs from our sample have
the low frequency break predicted in the simple qualitative picture
outlined in the Introduction. 
Fig.~\ref{fig:fbreak_forb} shows the dependence of the  best fit value
of the break frequency on the orbital frequency of the binary system.
There is an obvious positive correlation between these two frequencies
broadly consistent with eq.(\ref{eq:fvisc2forb}). 
Remarkably, this correlation holds over 3 order of magnitudes in the
orbital frequency, from compact  binaries 4U~1820-303 and 1H1627--673 with 
$P_{\rm orb}\approx 12$ and $\sim 40$ min to extremely wide  system 
GRS~1915+105 with $P_{\rm orb}\approx 33.5$  days.
The value of $f_{\rm break}/f_{\rm orb}$ ranges from $\sim 0.1-0.2$ for
long period binaries to $\sim 1-2$ for the more compact ones.
The broad band spectra shown in Fig.\ref{fig:pds_broadband} confirm
that the breaks are indeed unique and well defined features in the
power spectra. 

In the following we assume that the break frequency corresponds to the 
viscous time scale of the accretion disk, and they are related by a
simple linear  relation:  
\begin{equation}
t_{\rm visc}^{-1}=f_{\rm visc}= \beta^{-1} f_{\rm break}
\label{eq:tvisc_tbr}
\end{equation}
Note that this simplified picture does not take into account
complexity of the power density distribution near the viscous time
scale.

With this definition the break frequency is related to the orbital
frequency via:
\begin{equation}
\frac{f_{\rm break}}{f_{\rm orb}} = \beta \ \frac{3\pi \alpha}{\sqrt{1+q}} 
              \left(\frac{H_d}{R_d}\right)^2 
               \left(\frac{R_d}{a}\right)^{-3/2}
\label{eq:fbr2forb}
\end{equation}

\subsection{Cir~X-1}
\label{sec:cirx1}

Cir X-1 appears to deviate from the general trend in
Fig.\ref{fig:fbreak_forb}, having the break/viscous time scale by a
factor $\sim 10$ shorter than expected given its orbital period. 

The significant orbital modulation of the X-ray activity suggests a
highly eccentric orbit in this binary system \citep{murdin80}. 
\citet{johnston99} proposed a model, where the system
consists of a neutron star on a highly eccentric, $e\sim 0.7-0.9$
orbit around a $\sim 3-5$ $M_\odot$ sub-giant companion. In this model
the donor star fills its Roche lobe and the mass transfer occurs 
only during the periastron passage. Outside the periastron the donor
star is detached from its Roche lobe surface and the mass transfer in
the binary stops.

In such a system the disk radius would be defined, to first 
approximation, by the minimal separation between the components,
$a_{\rm min}=(1-e)\ a$. As the eq.(\ref{eq:fvisc2forb}) was derived
for a circular orbit, the following substitution should be made in
interpreting the Cir X-1 data: 
\begin{eqnarray}
a \rightarrow a\ (1-e) \hspace{1.2cm}\\
\nonumber
P_{\rm orb} \rightarrow P_{\rm orb}\ (1-e)^{3/2}
\label{eq:curx1}
\end{eqnarray}

Plotted in Fig.\ref{fig:fbreak_forb} are two point for Cir X-1.  The
open circle corresponds to the observed orbital period of the source,
the filled circle corresponds to the orbital period corrected for the
eccentricity of the orbit assuming $e=0.7$. 
With this substitution the consistency with other sources is
restored.

\subsection{Viscous time and disk thickness}
\label{sec:tvisc_h2r}

Although the break frequency does increase linearly with the binary
orbital frequency, the ratio $f_{\rm break}/f_{\rm orb}\sim 0.2-2$ is
notably larger than predicted for the Shakura-Sunyaev disk, 
$\left(f_{\rm visc}/f_{\rm orb}\right)_{\rm std} \sim 0.005-0.07$.
The latter range of values is shown as the hatched area in the 
Fig.~\ref{fig:fbreak_forb}.
Obviously, larger values of $f_{\rm visc}/f_{\rm orb}$ imply that the
disk viscous time as traced by the position of the break on the PDS is
by a factor of $\ga 5-10$ shorter than predicted by the theory.
As it follows from eq.(\ref{eq:fvisc2forb}) and (\ref{eq:fbr2forb}), 
there are several parameters affecting the disk viscous time, of which
the strongest dependence is upon the disk size and thickness.

\begin{figure}
\centering
\includegraphics[width=\columnwidth]{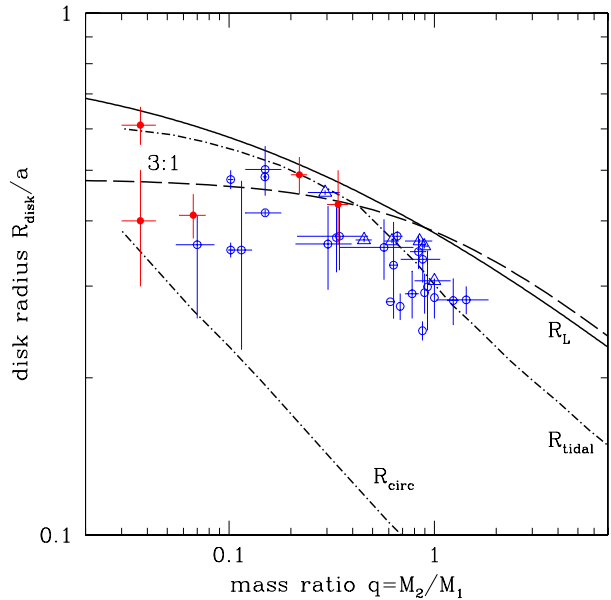}
\caption{Theoretical and observational constrains on the disk radius.
The solid curve shows the effective radius of the Roche lobe of the
primary. Two dash-dotted curves show the circularization
and tidal truncation radius. The dashed line is the radius of 3:1 tidal
resonance. The open circles and triangles are
measurements  and lower limits for  CVs in the outburst state. 
The small filled circles show the disk radius in LMXBs. 
The reference for these optical data are given in the text.
} 
\label{fig:rdisk}
\end{figure}

The theoretical and observational constrains on the disk size are
summarized in Fig.\ref{fig:rdisk}.
It has two well known theoretical limits. Due to the
angular momentum conservation, the disk radius can not be smaller than
the circularization radius.  The results of the numerical calculations
of \citet{lubow75} can be approximated by 
\begin{equation}
\frac{R_{\rm circ}}{a}=0.074\ \left ( \frac{1+q}{q^2} \right )^{0.24}
\end{equation}
This approximation is accurate to $\approx 3\%$ in the range of the
mass ratios $0.03\le q \le 10$. The upper limit on the disk size is
given by the tidal truncation radius \citep{pringle77, ichi94}. 
In the case of small pressure and viscosity its value is close to the
radius of the largest non-intersecting periodic orbit
\citep{pacz77}. The latter can be approximated by:
 \begin{equation}
\frac{R_{\rm tidal}}{a}=0.112+\frac{0.270}{1+q}+\frac{0.239}{(1+q)^2}
\end{equation}
accurate to $\approx 3\%$ in the range of $0.06\le q \le 10$.
These two limits on the disk radius are shown as thick dash-dotted
lines in Fig.\ref{fig:rdisk}. 
On the observational side, there is a number of spectroscopic and
photometric determinations 
of the disk radius in CVs \citep{hessman88, hessman90,
rutten92,harlaftis96, harrop96} and fewer in LMXBs \citep{orosz99,
shahbaz04, torres04, zurita00}. These data indicate
(Fig.\ref{fig:rdisk}) that for a steady
disk the disk radius is close to the tidal truncation radius as
defined by \citet{pacz77}, with possible exception of the small-q
systems (section \ref{sec:tvisc_lindblad}).

With these values of the disk size and accepting $\alpha\sim 0.5$ as a
plausible value of the viscosity parameter, the disk thickness of
\begin{equation}
H/R\ga 0.1\  \alpha_{0.5}^{-1/2} \beta^{-1/2}
\label{eq:h2r}
\end{equation}
is required in order explain observed values of the break frequency.

This conclusion relies on the association of the break in the power 
density spectra with the viscous time scale of the disk and  as long
as the assumption $\beta=f_{\rm break}/f_{\rm visc}\sim 1$ 
is valid, it is very robust. 
Indeed, in order to describe the data with 
$H/R\sim {\rm few\ \times\ } 10^{-2}$, as  predicted by  the standard
theory, one would need to  increase dramatically the $\alpha$-parameter,
up to implausibly high values of $\alpha\sim 5-800$. 
Alternatively, short viscous time of the
disk could be achieved by reducing the Roche-lobe filling factor of the
disk, down to $R_d/a\sim 0.005-0.05$. This is significantly smaller
than the circularization radius and contradicts to both theoretical
expectations and observations of the disks in CVs and LMXBs
(Fig.\ref{fig:rdisk}). 

The latter possibility can be also interpreted that only the very
inner part of the accretion disk, $R/a\sim 0.005-0.05$, contributes
to the observed variability of LMXBs in X-rays. This  can not be
excluded a priori. We note however, that in this case the sharpness
of the breaks observed in the power spectra of at least some sources 
(e.g. Sco X-1, EXO 0748--676, etc) implies that the transition from
the (outer) region of negligibly small amplitude of the $\dot{M}$
perturbation to the (inner) region of  relatively large ones should
occur in a rather narrow range of radii.

\begin{figure*}
\centering
\includegraphics[width=\hsize, bb=40 266 580 540]{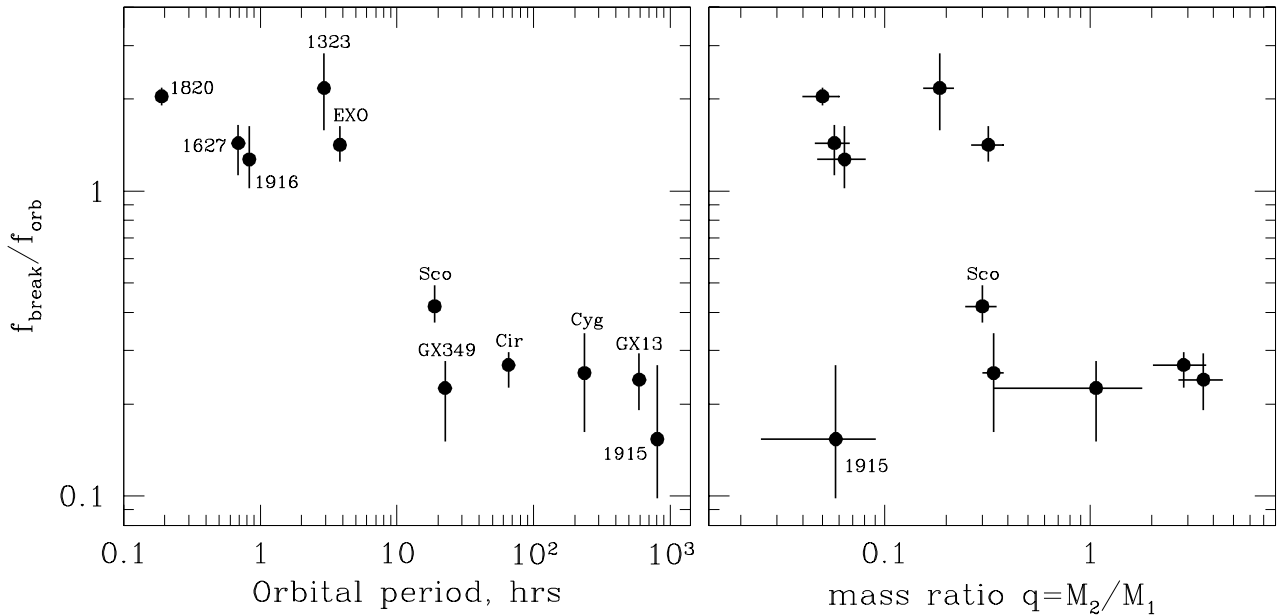}
\caption{Dependence of the ratio $f_{\rm break}/f_{\rm orb}$ on the
orbital period of the binary system ({\em left}) and its mass ratio
({\em right}).} 
\label{fig:fbr2forb}
\end{figure*}

\subsection{Viscous time in wide and compact systems}
\label{sec:tvisc_lindblad}

As is evident from Fig.\ref{fig:fbreak_forb}, there is a significant
dichotomy in the value of $f_{\rm break}/f_{\rm orb}$ ratio between
long- and short-period binaries. 
This is further illustrated by Fig.\ref{fig:fbr2forb}
where the ratio $f_{\rm break}/f_{\rm orb}$ is plotted against the
binary orbital period and the mass ratio. The average values of 
$f_{\rm break}/f_{\rm orb}$ are $\approx 0.23$ and $\approx 1.7$ for
the wide and compact systems correspondingly. The sense of the
difference is that the compact systems 
have systematically shorter, by a factor of $\sim 8$, viscous time
expressed in the orbital periods than the wide ones, the boundary
lying at $P_{\rm orb}\sim 10$ hours or $q\sim 0.3$.
The fact that this bimodality occurs between compact and wide systems
or, equivalently, (with the exception of GRS1915+106) between small-
and large-q binaries suggests that it may be caused by the excitation 
of tidal resonances in the accretion disk.

The phenomenon of tidal resonances in accretion disks is well studied
in the context of dwarf novae \citep{whitehurst88, whitehurst91,
lubow91}. 
The resonance occurs if the angular frequency of the orbital motion of
the particle  in the disk is commensurate with the angular frequency of
the orbital motion of the secondary. For a Keplerian disk, the
location of the resonant orbits is given by 
\begin{equation} 
\frac{R_{\rm res}}{a}=\left( \frac{k}{m} \right)^{2/3}\,(1+q)^{-1/3}
\end{equation} 
where $k,m$ are integers \citep[e.g.][]{whitehurst91}. An obvious
condition for a resonance being excited is that this radius lies
within the accretion disk. From analytical studies and numerical
simulations it is known that the strongest resonance occurs at the
lowest order commensurability, 
$\Omega_{\rm K}:\Omega_{\rm orb}=2:1$
\citep[e.g.][]{whitehurst91}. 
However the 2:1 resonant radius lies within the accretion disk only at
extreme values of the mass ratio, $q\la 10^{-2}$. The next strongest
resonance, most important in the context of binaries with not too
extreme values of the mass ratio lies  at the 3:1
commensurability. The higher order resonances are probably not excited
in the accretion disks. 
The dependence of the 3:1 resonance radius on the mass ratio $q$ is
shown by the dashed line in Fig.\ref{fig:rdisk}, confirming the well
known fact that the $3:1$ resonance can be excited in the systems with
the mass ration $q\la 0.3-0.4$. Note that the precise value of the
threshold $q$ depends on the definition of the tidal truncation
radius, due to  small angle between the two curves in
Fig.\ref{fig:rdisk}.  
Under the assumption  that $R_{\rm tidal}=0.9 R_L$ it is $q\approx
0.33$. With the definition of $R_{\rm tidal}$ used in
Fig.\ref{fig:rdisk} the threshold value is slightly larger.

Numerical simulations of the non-linear stage of the instability have
shown that excitation of tidal resonances results in a significant
asymmetry of the accretion disk and causes its precession in the
reference frame of the binary \citep{whitehurst88, whitehurst91,
lubow91}. Additionally, the accretion disk is truncated near the
resonant radius. This explains the values of the disk radii smaller than
the tidal truncation radius observed in CVs and LMXBs with small mass
ratios (Fig.\ref{fig:rdisk}).

To conclude, with exception of GRS1915+105, discussed in section
\ref{sec:grs1915}, the excitation of the 3:1 
inner Lindblad resonance provides a natural explanation of the
dichotomy in the viscous time scale between wide and compact systems.

\section{Discussion.} 
\label{sec:discussion}

\subsection{Thickness of the disk}

\subsubsection{Semi-thick disk or coronal flow?}
\label{sec:corona}

The values of the viscous time scale of the disk inferred by the
position of the low frequency break in the power spectra of LMXBs
require rather large disk thickness, 
$H/R\ga 0.1\  \alpha_{0.5}^{-1/2} \beta^{-1/2}$, 
by a factor of $\ga 3-5$ exceeding the prediction of the standard
theory of the accretion disks. 
Taken at the face value, $H/R\ga 0.1$ appears to agree
with the statistics of the eclipsing systems among LMXBs 
\citep[e.g.][]{milgrom78, narayan04}.
However, large values of $H/R$ result in small optical depth of the
outer disk, which is inconsistent with numerous observations 
of the optical emission from  accretion disks, as discussed below. 

\begin{figure}
\centering
\includegraphics[width=\hsize]{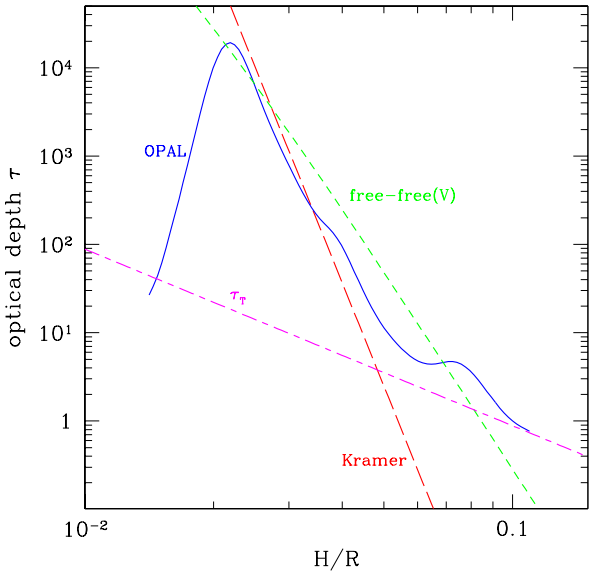}
\caption{Dependence of optical depth of the disk at $R=5\cdot 10^{10}$
cm on the $H/R$ ratio for $\dot{M}=5\cdot 10^{17}$ g/s, $M=1.4$
$M_\odot$, $\alpha=0.5$.
The solid line shows the optical depth based on the OPAL opacities
\citep{opal}, long dashed line -- Kramer law 
($k_R=5\cdot 10^{24}\,\rho\, T^{-7/2}$ cm$^2$/g), short dashed --
free-free absorption in the optical V-band, dash-dotted -- Thompson
optical depth. 
} 
\label{fig:tau}
\end{figure}

We consider  the  structure of an accretion disk with some value 
of the vertical scale-height $H/R\sim 0.1$, 
without specifying the source of the additional energy dissipation.  
Under the assumption of vertical hydrostatic equilibrium and
stationarity, both the mid-plane disk temperature and the density are
steep functions of the  disk thickness:
$T/T_{\rm vir}\propto (H/R)^2$, 
$\rho\propto (H/R)^{-3}$. 
Due to strong dependence of the Rosseland mean opacity on the 
density and temperature, typically as
$k_R\propto \rho^{\approx 1}\, T^{\approx -3.5}$ in the parameters
range of interest, the optical depth of the disk $\tau_R=k_R\rho H$ 
is also steeply decreasing function of the disk thickness,
$\tau_R\propto (H/R)^{-\gamma}$ with the power law index $\gamma$
changing from $\sim 12$ to $\sim 3$ as the $H/R$ increases
(Fig.\ref{fig:tau}).
Therefore, even a moderate increase of the disk thickness would lead to 
dramatic decrease of its optical depth. 
As a result, for $H/R\sim 0.1$, the accretion disk in the majority of
LMXBs from our sample would be optically thin. This would contradict
to a number of observational facts, mostly from the optical band,
proving existence of the optically thick outer disk in  LMXBs.
Indeed, the free-free optical depth in the optical V-band equals 
\begin{eqnarray}
\tau_{ff}({\rm V-band})\approx            \hspace{5.5cm}         \\
\nonumber
3.8\cdot 10^{-2}
\ln\left(\frac{T}{5500\rm~K}\right)
\dot{M}_{17}^2\, M_{1.4}^{-5/2}\, \alpha_{0.5}^{-2}\, 
\left(\frac{H}{R}\right)_{0.1}^{-8} R_{10}^{-1/2}
\end{eqnarray}
where $(H/R)_{0.1}$ is the disk geometrical thickness normalized to
0.1. The above approximation is
accurate to $\sim 10\%$ in the  $10^4\le T \le 10^8$ K temperature
range, the logarithmic term due to the Gaunt factor varies from 2.9 to
7.5 for the temperature in the range $10^5-10^7$ K. 
For the parameter of  binaries from our sample the optical depth in
the V-band would be $\tau_{ff}\ll 1$, with the exception of sources
with the largest $\dot{M}$ (Sco X-1, GX349+2 and 4U1820) 
where $\tau_{ff}\sim 1-10$. The low optical depth in the V-band would
conflict with numerous spectroscopic observations of the optical
emission from the LMXBs, suggesting existence of the continuum
emission originating from the optically thick outer accretion disk
\citep[e.g.][]{hynes02,chaty03}. 
On this basis the possibility of the semi-thick, $H/R\sim 0.1$
outer disk in LMXBs can be excluded.

Instead, one can envisage a two
phase accretion flow with a Shakura-Sunyaev-like disk in the mid-plane
surrounded by the tenuous optically thin coronal flow with the aspect
ratio $H/R\sim 0.1-0.2$ and temperature of 
$T\sim 10^{-2}\, T_{\rm vir}$. 
In order to explain the prominence of the low frequency break in the power
spectrum,  a significant fraction of the accretion has to take place
in this corona 
\begin{equation}
\dot{M}_{\rm corona}\sim \dot{M}_{\rm disk}
\end{equation}
The temperature and density in the coronal flow are:
\begin{eqnarray}
T\approx 1.4\cdot 10^5 \ M_{1.4}\,
\left(\frac{H}{R}\right)_{0.1}^2 R_{11}^{-1}
{\rm ~~[K]}
\end{eqnarray}
\begin{eqnarray}
n\approx 3.6\cdot 10^{13} \ f_c\, \dot{M}_{17}\, M_{1.4}^{-1/2}\,
\alpha_{0.5}^{-1}
\left(\frac{H}{R}\right)_{0.1}^{-3} R_{11}^{-3/2}
{\rm ~~[cm^{-3}]}
\end{eqnarray}
where $f_c=\dot{M}_{\rm corona}/\dot{M}$.  As the surface mass density
depends on the thickness of the disk as $\Sigma\propto (H/R)^{-2}$,
such a corona will contain a small fraction of the mass of the
accreting matter:
\begin{eqnarray}
\frac{\Sigma_{\rm c}}{\Sigma_{\rm d}}\sim \frac{f_c}{1-f_c}\, 
\left(\frac{H_c}{H_d}\right)^{-2}
\la 10^{-1} 
\end{eqnarray}
assuming $fc\sim 0.5$.

The vertical and radial column density of the coronal flow are $[cm^{-2}]$:
\begin{eqnarray}
\nonumber
n\,H\approx 3.6\cdot 10^{23} \ f_c\, \dot{M}_{17}\, M_{1.4}^{-1/2}\,
\alpha_{0.5}^{-1}
\left(\frac{H}{R}\right)_{0.1}^{-2} R_{11}^{-1/2}
\\
n\,R\approx 3.6\cdot 10^{24} \ f_c\, \dot{M}_{17}\, M_{1.4}^{-1/2}\,
\alpha_{0.5}^{-1}
\left(\frac{H}{R}\right)_{0.1}^{-3} R_{11}^{-1/2}
\label{eq:corona-nh}
\end{eqnarray}

The plausible mechanism of formation of the coronal flow is the disk
evaporation process originally proposed by \citet{meyer94} for
dwarf novae.
In X-ray binaries the physics of the disk evaporation is significantly
affected by the illumination and Compton heating and cooling of the
disk-corona system by the X-ray emission produced in the vicinity of
the relativistic object. Although the the self-consistent treatment of
the problem is yet to be done, the likely net effect of the X-ray 
irradiation is to increase the fraction of $\dot{M}$ in the coronal flow  
\citep[][F.Meyer, private communication]{meyer00, jimenez02}.

\subsubsection{Other evidence of the coronal flow}

An independent evidence of the existence of a diffuse
ionized  gas above the accretion disk plane is provided by
observations of LMXBs with high inclination -- the ADC sources and
dippers. Based  on observations of partial X-ray eclipses in
4U1822--371, \citet{white81} concluded that the compact X-ray source
in this system is diffused by a large moderately Compton thick highly
ionized corona located above and below the accretion disk.
Detailed modeling of the eclipse light curves in the ADC
sources 4U1822--371 and 4U2129+47 \citep{white82} yielded estimates of
the radial extent of the accretion disk corona in these sources, 
$R_c\sim (1-2)\cdot 10^{10}$ cm and $R_c\sim (3-6)\cdot 10^{10}$.
These numbers are comparable with the value of the tidal radius in the
systems, $R_{\rm tidal}\sim 3.9\cdot 10^{10}$ cm. The partial nature of
the eclipses in these sources, with the eclipsed flux at the level of
$\sim 50\%$ of the uneclipsed value indicated that the corona had
non-negligible radial optical depth, $\tau_T\sim 1$, in agreement with
the estimate of eq.(\ref{eq:corona-nh}).

High resolution spectroscopic observations of LMXBs with Chandra
and XMM-Newton gratings are revealing complex absorption/emission
features in their X-ray spectra \citep{cot01,jimenez03, kallman03,
boirin04}.  These features are mostly pronounced in the systems with
high binary inclination angle and  suggest the presence of tenuous 
photo-ionized plasma orbiting the compact object above the orbital
plane of the binary. 

From the analysis of the recombination  emission lines of H- and
He-like ions of O, Ne and Mg in  EXO0748--676
\citet{jimenez03} constrained the density and radial extend of the
photoionized  plasma, $N_e\sim 10^{12}-10^{15}$ cm$^{-3}$, 
$7\cdot 10^{9}\la R \la 6\cdot 10^9$ cm and the temperature $T\la
2\cdot 10^5$ K. This range of the parameters is consistent with the
numbers computed from the above formulae for 
EXO0748--676 and assuming the aspect ratio of the coronal flow 
$H/R=0.1$: the tidal truncation radius $R_{\rm tidal}\sim 4.8\cdot
10^{10}$ cm, density in the coronal flow $N\sim 2.7\cdot 10^{13}f_c$
cm$^{-3}$, and temperature of  $T\sim 3\cdot 10^5$ K.

Based on the observations of the narrow absorption lines of the H- and
He-like Fe in the spectrum of 4U1916--053, \citet{boirin04} estimated
the ionization parameter and the column density of the 
absorbing gas: $\log(\xi)\approx 3.9$, $NH\sim 2\cdot 10^{22}$
cm$^{-2}$. From these numbers one can estimate the radial extent and the
density of the absorbing gas, $R\sim 1.6\cdot 10^{10}$ cm, $N\sim
1.3\cdot 10^{12}$ cm$^{-3}$. The tidal radius in this system is
$\approx 2\cdot 10^{10}$ cm, in agreement with the value derived from
the XMM-Newton data. The density in the coronal flow with $H/R=0.1$ is 
$N\sim 6.7\cdot 10^{12}f_c$ cm$^{-3}$. The latter number is somewhat
larger than derived from the XMM-Newton data. This can be explained by
smaller inclination angle in this source resulting in lower density
along the line of sight. Indeed,  unlike  EXO0748--676 showing deep
X-ray eclipses, 4U1916--053 shows only X-ray dips. 

\subsubsection{$\dot{M}$ variations due to geometrically thin disk}

In this picture, the $\dot{M}$ variations in the geometrically thin
Shakura--Sunyaev disk should give rise to the second very low
frequency power law component, revealing itself below the break frequency. 
Such power law component is indeed observed in the power spectra of
some of the sources. Most obvious its presence is in Cir X--1 and
4U1916--053 and, to less extent, in Sco X--1
(Fig.\ref{fig:pds_broadband}).  Remarkably, the slope of the power
spectrum before and after the break are very similar, suggesting
similar nature of the processes causing the variability.
Furthermore, the second break should be
expected at the low frequencies, corresponding to the (larger) viscous 
time of the geometrically thin disk. The frequency of the second
break is described by eq.(\ref{eq:fbr2forb}) with the $H/R$ given by
eq.(\ref{eq:h2r_ss73}). An evidence of the second break may be seen in
the power spectra of Sco X-1 and, possibly, of 4U1916--053.
In the former case the break frequency equals
$f_{\rm break}^{(2)}= 8_{-4}^{+7}\cdot 10^{-8}$ Hz and the ratio
$f_{\rm break}^{(2)}/f_{\rm orb}=5_{-3}^{+5}\cdot 10^{-3}$. This value
is within the range predicted for standard Shakura--Sunyaev disk
with the aspect ratio of $H/R\sim {\rm few}\cdot 10^{-2}$, 
eq.(\ref{eq:fvisc2vorb_std}).
The value of the break frequency in the case of 4U1916--053 
is probably $f_{\rm break}^{(2)}\sim 10^{-7}$ Hz, although this number
is less secure than that for Sco X-1. 
If correct, this is significantly, by $\sim 2-3$ dex smaller that
expected in the above scenario.

\bigskip
\bigskip

In the conclusion of this section we note that due to 
power law dependence of the viscous time scale on the disk radius,
$t_{\rm visc}\propto r^{3/2}$, the radial drift velocity of the matter
in the accretion flow  should be increased throughout the most 
($\sim 80-90\%$) of the radial extent of the accretion disk. Therefore
the bulge at the outer edge of the disk, resulting from the
disk--stream interaction \citep[e.g.][]{mason89} is insufficient as it
is located at the outer edge of the disk. 
For the same reason this analysis based on the viscous time of the
disk probes the outer disk and is insensitive to the conditions in its
innermost parts, e.g. $R\la 10^4 R_g$,  which contribution to the
total viscous time is negligibly small.

\subsection{Tidal resonances and viscous time scale of the disk}

The dichotomy in the viscous time between long- and short-period
systems is naturally explained by excitation of the 3:1 inner Lindblad 
resonance. This suggestion is supported by detection of superhumps in
several LMXBs with small mass ratios
\citep[e.g.][]{haswell01,zurita02}. Importantly, the superhumps are
detected not only in transient sources but also in one or two
persistent LMXBs, including one source from our sample, 4U1916--053
\citep{callanan95}. 

The specific mechanism by which the tidal instability is affecting the 
viscous time of the disk  is, however, unclear. It is also unclear if
the coronal flow in the small-q system has indeed $H/R\sim 0.3$ and
$T\sim 0.1\, T_{\rm vir}$ as it is formally required by the observed 
$t_{\rm   visc}\sim P_{\rm orb}$.
Several possibilities can be mentioned:
\begin{enumerate}
\item 
mass transfer in tidal waves excited in the disk by the tidal forces. 
It has been shown theoretically  and confirmed in numerical
simulations, that in the non-linear regime strong tidal waves are
excited in the accretion disk \citep{spruit87,lubow91,truss02}. 
These waves, propagating in the disk with the speed significantly
exceeding the  radial drift velocity of the matter,
can  decrease the effective viscous time scale.
If correct, the aspect ratio and coronal temperature in the small-q
systems may be not significantly different from those in the large-q 
ones. 

\item
heating of the outer disk due to resonant tidal interaction
\citep{whitehurst91,truss02} can increase the  temperature of the disk
and/or coronal flow and, correspondingly, decrease the viscous time
scale. As was mentioned above, in order to explain the observed
viscous time scale  the coronal flow temperature of 
$T\sim 0.1\, T_{\rm vir}$ is required.

\item non-trivial definition of the viscous time for an eccentric and
precessing disk. In the fully developed instability the particles in
the disk move on non-circular trajectories with large
eccentricity. Therefore, the average velocity of the radial drift of
the disk particles can be significantly different from the value
predicted by the standard disk theory for a circular Keplerian disk. 
\end{enumerate}

We note, that the truncation of the accretion disk at the 3:1
resonance radius (Fig.\ref{fig:rdisk}) can be excluded as the cause of
the shorter viscous time scale in the small-$q$ systems, because the
change of the outer disk radius by a $\la 20-30\%$ is insufficient
to explain the observed decrease of the viscous time by the factor of
$\sim 10$.

\subsubsection{Case of GRS1915+105}
\label{sec:grs1915}

GRS1915+105 clearly stands out among other 
small-q systems (Fig.\ref{fig:fbr2forb}, right panel). 
Although it has small mass ratio, $q=0.058\pm 0.033$ \citep{grs1915},
its ratio $f_{\rm break}/f_{\rm orb}$ is close to the value found in
the binaries with $q\ga 0.3$. 
The most plausible explanation of such behavior is related to the
transient nature of the source. 
We outline two possibilities:
\begin{enumerate}
\item the tidal instability is known to have a rather long growth time,
of the order of $\ga 10^2$ binary orbital periods 
\citep[e.g.][]{whitehurst88}.  
For GRS1915+105 $P_{\rm orb}\approx 33.5$ days and
$10^2 P_{\rm orb}$ corresponds to $\sim 10$ years. 
Therefore the time passed since the onset of the outburst ($\sim 10$
years) was insufficient for the tidal instability to fully develop.
In this scenario one would expect an increase of the break frequency
in the course of the outburst, unless the instability growth time is
significantly longer than $\sim 10$ years. 
However, no statistically significant
difference in the break frequency has been found between the first and
second halves of the data. This fact speaks against the above
suggestion. 
\item
due to the large size of the system ($\sim 8\cdot 10^{12}$ cm), the
irradiation effects are insufficient to ionize the outer disk in GRS1915+105
which  is in the cold, low-viscosity state (H.Ritter,
private communication). Therefore its outer boundary is located near
the circularization radius rather than has expanded to the tidal
radius, as is the case for the persistent sources. Therefore there is
no accretion disk at the  3:1 resonance radius
(Fig.\ref{fig:rdisk}) and  the tidal instability is not
excited. In this respect the present outburst of GRS1915+105 might be 
similar to normal outbursts in dwarf novae. As well known, for the
same reason the superhumps in dwarf novae are observed (tidal
instability is excited) only in during the superoutbursts but not
during the normal outbursts \cite[e.g.][]{osaki03}. 
\end{enumerate}

\subsection{Broad low frequency QPO features}
\label{sec:qpo}

For several sources broad QPO-like features are obvious in the power
density spectra. Most apparent these features are in the case of Cyg
X-2 and GRS1915+105 (Fig.\ref{fig:pds_asm}, \ref{fig:pds_exosat}). 
Among  plausible mechanisms of appearance of quasi periodic
variability in X-ray binaries on these time scales are the
radiation-driven warping and precession of the  accretion
disk \citep*[e.g.][]{maloney96} and mass-flow oscillations caused by diffusion
instability \citep{meyer90}. 
Both types of oscillations are caused by the irradiation of the
accretion disk by the primary and their characteristic time scale is
of the order of the viscous time scale of the disk. Accordingly, these
features should appear near the break frequency in the power density
spectrum. 

Another possible mechanism of appearance of broad low frequency QPOs 
near the viscous frequency of the disk was suggested by
\citet*{vikhl94}. This mechanism is related  to the anti-correlation of
the $\dot{M}$ perturbations in the accretion flow on the time scales 
$t\sim t_{\rm visc}$. The anti-correlation appears in the presence of
the steady mass supply through the outer boundary of the accretion
flow and is a direct consequence of the energy conservation law. In
this respect the broad QPO features observed near the viscous time
scale of 
the outer disk might be similar to the broad QPOs detected near  the
high frequency break of power density spectrum of Cyg X-1 in the hard
state, at frequency of $\sim 0.04-0.07$ Hz \citep{vikhl94_cygx1}. In
the latter case the values of the break frequency and of the QPO
centroid frequency are defined by the viscous time scale of the inner
hot flow, $R\la 10^2$ R$_g$, responsible for the production of the hard
Comptonized emission in the low spectral state of black hole
candidates.

\section{Summary} 
\label{sec:summary}
 
We studied the low frequency variability of low mass X-ray binaries in
order to search for the signatures in their power density spectra
related to the viscous time scale of the disk.
Our results can be summarized as follows:

\begin{enumerate}

\item 
As the viscous time is the longest time scale of the accretion
disk, the $\dot{M}$ variations produced in the disk due to various
instabilities should become independent of each other on the time
scales longer than the viscous time, i.e. 
$\left< \dot{M}(t)\dot{M}(t+\tau) \right> \rightarrow 0$ at 
$\tau\ga t_{\rm visc}$
Correspondingly, the power density of $\dot{M}$ perturbations produced in
the disk and of the observed  X-ray flux variations should become
independent of the frequency at $f\la t_{\rm visc}^{-1}$. 

\item
Using archival data of RXTE/ASM  and EXOSAT/ME we studied X-ray
variability of persistent LMXBs in the $f\sim 10^{-8}-10^{-1}$ Hz
frequency range.  In power density spectra of 11
sources out of 12 satisfying our selection criteria, we found a very
low frequency break.  The spectra approximately follow
$P_\nu\propto \nu^{-1.3}$ power law above the 
break and are nearly flat below the break 
(Figs. \ref{fig:pds_asm}--\ref{fig:pds_broadband}).
In some cases (Sco X-1, Cir X-1 and 4U1916--053) a second power law
component with the same slope appears at $f\ll f_{\rm break}$.  

\item
In a very broad range of  binary periods, from $P_{\rm orb}\sim 11$
min in ultracompact binary 4U1820-303  to $P_{\rm orb}\sim 33.5$ days
in GRS1915+105, the break frequency correlates with the orbital
frequency $f_{\rm orb}=1/P_{\rm orb}$ of the binary system 
(Fig.\ref{fig:fbreak_forb}), in good agreement with the theoretical
prediction for the viscous frequency  
$f_{\rm visc}=t_{\rm visc}^{-1}$
(eq.\ref{eq:fvisc2forb})

\item
Assuming that the low frequency break is associated with  the
viscous time scale of the disk, $f_{\rm break}\sim 1/t_{\rm visc}$,
the measured values of the break frequency imply that the
viscous time of the disk is  by a factor of $\ga 10$
shorter than predicted by the standard theory 
(section \ref{sec:tvisc_h2r}, Fig.\ref{fig:fbreak_forb}). 
Taken at the face value, this requires the relative height of the
outer disk $H/R\ga 0.1$.  Motivated by the low vertical optical depth of
such semi-thick accretion flow we propose instead, that 
significant fraction of the accretion occurs through the coronal flow
above the standard geometrically thin Shakura-Sunyaev disk  (section
\ref{sec:corona}, Fig.\ref{fig:tau}).  
The aspect ratio of the coronal flow, $H/R\ga 0.1$, corresponds
to the gas temperature of $T\ga 0.01 \, T_{\rm vir}$.  The corona has
moderate optical depth in the radial direction, $\tau_T\sim 1$, and
contains $\la 10\%$ of the total mass of the accreting matter. 
The estimates of the temperature and density of the corona are in
quantitative agreement with the parameters inferred by the X-ray
spectroscopic observations by Chandra and XMM-Newton of complex
absorption/emission features in the LMXBs with large inclination
angle.  

\item
We found a clear dichotomy in the $f_{\rm break}/f_{\rm orb}$  between
wide and compact system, the accretion flow in the compact systems
having $\sim 10$ times shorter viscous time expressed in the orbital
periods of the binary, than in the wide ones
(Fig.\ref{fig:fbr2forb}). The jump in 
the  $t_{\rm visc}/P_{\rm orb}$ ratio occurs at the binary system
mass ratio of  $q\sim 0.3$.
This strongly suggests that the dichotomy between small-q
and large-q systems is caused by the excitation of the 3:1 Lindblad
resonance in the small-q systems (section \ref{sec:tvisc_lindblad}).

\end{enumerate}

\section{Acknowledgments}
We are grateful to Friedrich Meyer and Hans Ritter for
numerous discussions of the physics of the accretion 
disks and superhumps phenomenon in dwarf novae.
VA would like to acknowledge the partial support from the 
President of RF grant  NS-2083.2003.2 and from the RBRF grant
03-02-17286. 
This research has made use of data obtained through the High Energy
Astrophysics Science Archive Research Center Online Service, provided 
by the NASA/Goddard Space Flight Center.

\appendix
\section{Power spectra from the the ASM data} 
\label{sec:method}
 
To calculate power density spectra from ASM light curves we used the 
method based on the auto-correlation function \citep[cf.][]{edelson88}. 
 
We begin with consideration of the equally spaced data. The measured
time series is $x_k$ and equals the number of counts detected in the
$k$-th time bin corresponding to the time interval of $t=(k-1)\Delta t
\div k\Delta t$,  where index $k$ runs from 0 to $N-1$, N is assumed to
be even. The discrete Fourier transform of the time series is defined as :
\begin{equation}
\hat{x}_j=\sum^{N-1}_{k=0}\, x_k\, \exp \left(i \frac{2\pi j k}{N}
\right)
\label{eq:ft}
\end{equation}
where $j$ changes in the range $1\div N/2$. The power density in the
$j$-th frequency bin, which center is $f_j=j/T$ Hz, expressed in units
of rms$^2$/Hz (rms  is fractional rms of variability) equals: 
\begin{equation}
P_j=\frac{2 |\hat{x}_j|^2}{N_{\rm ph}}\, \frac{1}{\left< r \right>}=
\frac{2 |\hat{x}_j|^2}{\left< r \right>^2T}
\label{eq:ft_power}
\end{equation}
where $N_{\rm ph}=\sum^{N-1}_{k=0}\, x_k$ is the total number of
counts in the time series, $T=N\Delta t$ is its total duration
and $\left< r \right>=N_{\rm ph}/T$ is the average count rate.

Combining eqs.\ref{eq:ft} and \ref{eq:ft_power} one finds
\begin{equation}
P_j=\frac{2}{\left< r \right>^2T}
\left( A_0\, N +
2 \sum_{l=1}^{N-1} (N-l) \left<A_l\right > \cos \frac{2\pi j l}{N}
\right)
\label{eq:acf_power}
\end{equation}
where $\left<A_l\right >$ is discrete auto-correlation function:
\begin{equation}
\left<A_l\right >=\left<x_k\, x_{k+l}\right >
\end{equation}
The eq.\ref{eq:acf_power} expresses the well known fact that power
equals the cosine transform of the auto-correlation function.
The noise level in the power density spectrum equals:
\begin{equation}
P_{\rm noise}=2\,\frac{\sum_{k=0}^{N-1}\sigma_k^2}{\left< r \right>^2T}
\end{equation}
where $\sigma_k$ is the error of $x_k$. For Poisson errors in
$x_k$,  $\sigma_k^2=x_k$ and $P_{\rm noise}=2/\left< r \right>$, in
agreement with the conventional form \citep[e.g.][]{leahy83}.

The ASM time series consists of the count rate measurements $r_k$,
[cnts/sec],  at unevenly spaced moments $t_k$, $k=1,...,N_{\rm ASM}$. 
The values of $r_k$ are
averaged over the time interval ($\sim 90$ sec) much shorter than the
distance between adjacent bins ($\sim 90$ min). As before, the 
auto-correlation function $A_l$ is defined on the grid of $N$ time
bins, with the $k$-th bin corresponding to the time interval
$t=(k-1)\Delta t \div k\Delta t$, 
where index $k$ runs from 0 to $N-1$. 
The $\left<A_l\right >$ is computed as 
\begin{equation}
\left<A_l\right >=\left<(r_i-\left<r\right>)
(r_j-\left<r\right>)\right>\, 
\Delta t^2
\label{eq:acf_asm}
\end{equation}
where the averaging is performed  over all pairs $(i,j)$, $i>j$, for
which $\Delta t_{ij}=t_i-t_j$ falls in the $l$-th time bin of the
grid. 

The time interval $\Delta t$ and
the duration of the time series $T=N\Delta t$ are parameters of the
transformation and define the frequency range covered by the output
power density spectrum. Their choice depends on the effective Nyquist
frequency of the time series and its duration. Appropriate for the
parameters of the ASM light-curves are the following values:
$\Delta t\sim 5\cdot 10^3 \div 10^4$ sec 
and  $T\sim 10^7\div 10^8$ sec. 
Advantage of this method is that the correctness of the chosen values
of  $\Delta t$ and $T=N\Delta t$ can be easily checked using the
number of pairs $(i,j)$ used to compute the autocorrelation 
value  at the given time scale $l\Delta t$ and the uniformity of these numbers
across the full range of the time scales, from $\Delta t$ to
$T=N\Delta t$.
 
\begin{figure}
\centering
\includegraphics[width=\hsize]{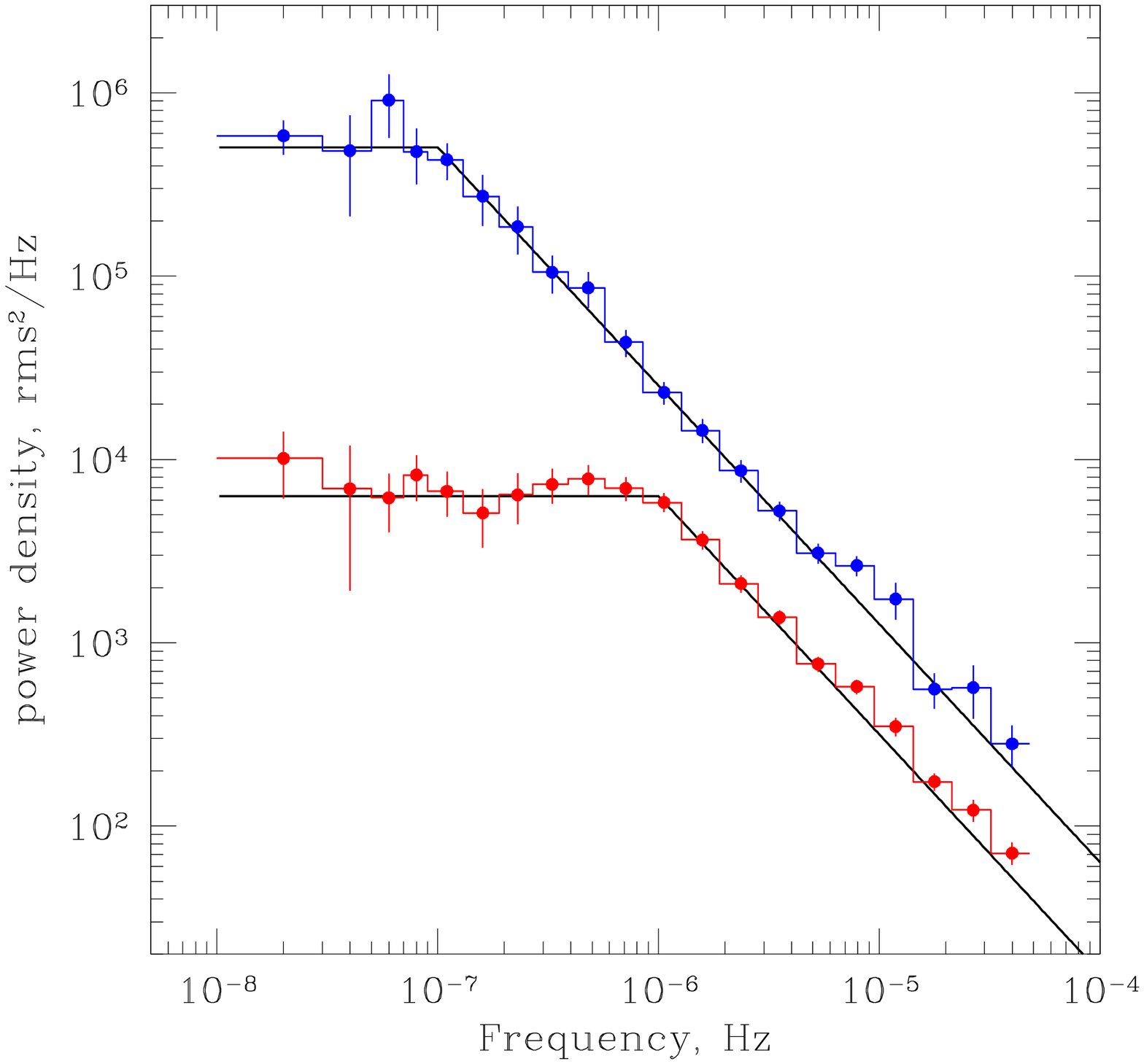}
\caption{Results of simulations. The power density spectra of
ASM light-curves  simulated as described in the text.
The solid lines show the model power density spectra used to generate
faked ASM light curves: power law $P_\nu=10^{-4} f^{-1.3}$ with low
frequency break at $10^{-6}$ and $10^{-7}$ Hz. The model with 
$f_{\rm break}=10^{-7}$ Hz  and corresponding  power density spectrum
are shifted upwards by the factor of 4 for clarity.
} 
\label{fig:pds_fake}
\end{figure}

With the auto-correlation computed according to eq.\ref{eq:acf_asm},
the power density spectrum can be easily computed using
eq.\ref{eq:acf_power}. As usual in the time series analysis, the
entire time series is divided into $n$ segments and the power spectrum
is computed for each segment separately. Their average and dispersion
give estimates of the average power density spectrum and its
uncertainty (under the assumption of stationarity).

If the errors for $r_k$ are known, the noise level in the output power
spectrum can be computed as 
\begin{equation}
P_{\rm noise}=2\,\frac{\sum_{k=1}^{N_{\rm ASM}}\sigma_k^2\,\Delta t}
{\left< r \right>^2 N_{\rm ASM}}
\label{eq:asm_noise}
\end{equation}
where $\sigma_k$ is the error of $r_k$ (in units of cnts/sec) and 
$N_{\rm ASM}$ is the number of ASM measurements used to compute one
power spectrum. In practice, the flux measurement errors given in the
ASM light-curves are slightly underestimated, therefore the noise level
computed with eq.\ref{eq:asm_noise} is somewhat imprecise
\citep{grimm02}.

The performance of the method described above was verified in
simulations. The initial light curve with the time resolution of $\sim
100$ sec was computed assuming the power law spectrum  with
slope of 1.3 and low frequency break at $f=10^{-6}$ Hz and 
$f=10^{-7}$ Hz and random Fourier phases. From this 
light curve the ASM time series was simulated using the measurements
times $t_k$ of the real ASM light curve of Cyg X-2. The obtained light
curve was randomized assuming Gaussian errors with the standard
deviation equal the real ASM measurement error for the given
measurement $r_k$. The obtained time series was analyzed with the same
code as used for the  analysis of the real ASM data with parameters 
$\Delta t=10^4$ sec and $T=5\cdot 10^7$ sec. The noise level 
estimated using eq.\ref{eq:asm_noise} was subtracted from the power
density spectra. 
The results of simulations are shown in Fig.\ref{fig:pds_fake} along
with the model power density spectra used to generate the initial
high resolution light curves.

\end{document}